\documentclass[12pt]{article}
\usepackage{epsfig}
\usepackage{amsmath}
\usepackage{amssymb}
\usepackage{pstricks,pst-plot}
\psset{unit=1cm,linewidth=1pt,arrowscale=1.5}
\usepackage[left=1in,right=1in,top=1in,bottom=1.25in]{geometry}
\usepackage{setspace}
\numberwithin{equation}{section}
\newcommand{\be}{\begin{equation}}
\newcommand{\ee}{\end{equation}}
\newcommand{\bea}{\begin{eqnarray}}
\newcommand{\eea}{\end{eqnarray}}
\newcommand\bra[1]{\left\langle #1 \right|}
\newcommand\ket[1]{\left| #1 \right\rangle}
\newcommand\half{\tfrac{1}{2}}
\newcommand\tr{\mathop{\rm tr}\nolimits}
\newcommand\str{\mathop{\rm str}\nolimits}
\newcommand\rar{C_{a/r}}
\newcommand{\CF}{{\cal F}}
\newcommand{\CA}{{\cal A}}
\newcommand{\HA}{{\hat A}}
\newcommand{\TZ}{{\tilde  Z}}
\newcommand{\Tz}{{\tilde  z}}

\begin{document}
\begin{titlepage}
\renewcommand\thefootnote{\fnsymbol{footnote}}

\begin{flushright}
TAUP--TH 2903/09\\
UTTG--02--10\\
arXiv:1003.2621v2\\
\end{flushright}

\begin{center}
{\Large\bf Searching for an Attractive Force\\ in Holographic Nuclear Physics}\\
\vrule height 25pt width 0pt
{\large Vadim Kaplunovsky}\footnote{vadim@physics.utexas.edu}\\
{\it Physics Theory Group, University of Texas\\
1 University Station, C1608, Austin, TX 78712, USA}\\
\vrule height 25pt width 0pt
{\large Jacob Sonnenschein}\footnote{cobi@post.tau.ac.il}\\
{\it School of Physics and Astronomy,\\
The Raymond and Beverly Sackler Faculty of Exact Sciences,\\
Tel Aviv University, Ramat Aviv, 69978, Israel.}\\
\end{center}

\vfil

\noindent
\centerline{\Large\bf Abstract}
\begingroup\singlespacing
We are looking for a holographic explanation of nuclear forces,
especially the attractive forces.
Recently, the repulsive hard core of a nucleon-nucleon potential was obtained
in the Sakai--Sugimoto model, and we show that a generalized version of that model
--- with an asymmetric configuration of the flavor D8 branes --- also has
an attractive potential.
While the repulsive potential stems from the Chern--Simons interactions of the
$U(2)$ flavor gauge fields in 5D, the attractive potential is due to a coupling of
the gauge fields to a  scalar field describing fluctuations of the flavor
branes' geometry.
At intermediate distances $r$ between baryons
--- smaller than $R_{\rm KK}=O(1)/M_{\omega\,\rm meson}$ but larger than the radius
$\rho\sim R_{\rm KK}/\sqrt{\lambda_{\rm 't\,Hooft}}$
of the  instanton at the core of a baryon
--- both the attractive and the repulsive potentials behave as $1/r^2$,
but the attractive potential is weaker:
Depending on the geometry of the flavor D8 branes, the ratio
$\rar=-V^{\rm attr}(r)/V^{\rm rep}(r)$ ranges from 0 to $\frac19$.
The 5D scalar fields also affect
the isovector tensor and spin-spin forces,
and the overall effect is similar to the isoscalar central forces,
$V(r)\to (1-\rar)\times V(r)$.
\par
At longer ranges $r\gtrsim R_{\rm KK}$, we find that
the attractive potential decays faster
than the repulsive potential, so the net potential is always repulsive.
This unrealistic behavior may be peculiar to the Sakai--Sugimoto-like models,
or it could be a general problem of the $N_c\to\infty$ limit inherent in holography.
\par\endgroup
\eject
\end{titlepage}
\renewcommand\thefootnote{\arabic{footnote}}
\setcounter{footnote}{0}

\begingroup
\parskip=0pt plus 0pt minus 0pt
\tableofcontents
\par\endgroup

%
\section{Introduction}
In recent years,
holography or gauge-gravity duality gave us a new approach to
hadronic physics (see \cite{Gubser:2009md} for a review).
It has been spectacularly successful at explaining many features
of the quark-gluon plasma such as its low viscosity \cite{Kovtun:2004de},
and there are some interesting results concerning the high-density
nuclear matter \cite{holnuc,BLL-RSVW}.
Motivated by this success, the authors wanted to apply holography
to one of the oldest problems of nuclear physics:
{\it The interactions between nucleons are very strong, so why
isn't the nuclear matter relativistic?}
Instead, the bulk binding energy of the nuclear matter is
only 1.7\% of $Mc^2$, about 16~MeV per nucleon. 

The usual explanation of this puzzle involves a near-cancellation
between the attractive and the repulsive nuclear forces: The
attractive potential is  only a little bit stronger than the repulsive
potential, and the difference is rather small.
For example, in the Walecka's mean-field model \cite{Walecka:1974qa},
the attractive potential due to $\sigma$-meson field is 400~MeV
while the repulsive potential due to $\omega$-meson field is 350~MeV;
there is also the Fermi motion energy of about 35~MeV/nucleon,
so the net binding energy is only 16~MeV/nucleon.
There have been many similar (but more elaborate) models since Walecka,
but they all beg the same question:
Why is the attractive $-(\overline\Psi\Psi)^2$ interaction between
nucleons only a little bit stronger than the repulsive
$+(\overline\Psi\gamma^\mu\Psi)^2$ interaction?
Is this a coincidence depending on quarks having precisely 3~colors
and the right masses for the $u$, $d$, and $s$ flavors?
Or is this a more robust feature of QCD that would persist for different $N_c$
and any quark masses (as long as two flavors are light enough)?

The most direct way of applying holography to these issues would be
to build a holographic model of the bulk nuclear matter.
Unfortunately, this approach is troubled by the large $N_c$ limit which
is inherent in the Holographic QCD.
Indeed,  even taking the leading $1/N_c$ corrections into account
is very hard in HQCD because it requires doing string loop calculations 
on the ``gravity'' side of the gauge-gravity duality.
But for large $N_c$, the low-temperature low-pressure phase of 
the bulk nuclear matter becomes a crystalline solid instead of
the Fermi liquid for the real-life $N_c=3$, and its other properties
--- such as density or the binding energy --- could also be quite different.

Paradoxically, direct holographic modeling works for exotic phases of nuclear matter
--- such as the quark-gluon plasma and maybe the high-density solid phase,
if it exists\footnote{%
	Similar to the liquid helium solidifying under pressure,
	the $N_c=3$ nuclear matter may also have a crystalline high-pressure phase.
	Although at very high pressures and densities, the nucleons are believed
	to merge into a quark liquid, and it is not clear if the nucleons form
	a lattice before merging, of if there is a direct transition from the nuclear liquid
	to	the quark liquid.
	If the solid nuclear phase exists at all and behaves like a semi-classical crystal,
	then its structure should not depend much on the $N_c$ and it could be modeled
	 using large--$N_c$ methods such as
	the  skyrmion lattices of \cite{Klebanov:1985qi},
	or the holographic instanton lattices of \cite{Rho:2009ym}.
	Alas, judging the phenomenological success of such models is rather difficult because
	the high-pressure nuclear matter is hard to study experimentally;
	the best data comes from modeling neutron stars, and we still do not know for
	sure if their interiors are solid or liquid.
	}
--- but not for the good old nuclei themselves.
However, we may still use holography to obtain phase-independent features of
nucleons and nuclear forces, but relating  those features to the experimental properties
of real nuclei has to be done by some other methods.
So in this article, instead of trying to model whole nuclei, we focus on a rather humble
problem of obtaining an attractive two-body nuclear force from the Holographic QCD.

The first holographic model of a baryon appeared in  \cite{Witten:1998xy} and
\cite{Gross:1998gk} in the ${\rm AdS}_5\times S^5$ context:
A D5 brane wrapping the $S^5$ had $N_c$ strings attached to it, while the
opposite ends of those strings connected to the $N_c$  external quarks
at the boundary of the ${\rm AdS}_5$ space.
Similar ``external'' baryons were constructed in confining backgrounds in
\cite{Brandhuber:1998xy}.
To make a baryon out of dynamical rather than external quarks one needs to add
$N_f$ flavor branes to the holographic model;
usually one takes $N_f\ll N_c$ so the flavor branes act as probes of the background
created by the color branes.
A prototypical model of this kind was constructed by Sakai and Sugimoto  \cite{Sakai:2004cn}:
Starting with the Witten's  model \cite{Witten:1998zw} of
$N_c$ D4--branes on a circle (with antiperiodic boundary conditions for the fermions
to break the ${\cal N}=4$ SUSY to ${\cal N}=0^*$),
they have added $N_f$ D8 and $N_f$ $\overline{\rm D8}$ branes.
On the gravity side of the duality, the 10D geometry is warped ${\bf R}^{1,3}
\times S^4\times{}$a cigar, while the D8 and $\overline{\rm D8}$ branes connect
to each other and span
${\bf R}^{1,3} \times S^4\times{}$a U-shaped line on the cigar
(see figure~(\ref{Ushape}) on page~\pageref{Ushape}).
A holographic baryon comprises a D4 brane wrapping a compact $S^4$ and $N_c$
open strings connecting this D4 to the flavor D8 branes.
To minimize the baryon's energy, the D4 brane acting as a baryonic vertex
becomes embedded in the D8 branes \cite{Seki:2008mu}, and for $N_f>1$
it dissolves into an instanton of the $U(N_f)$ gauge theory on the flavor branes.

Sakai, Sugimoto {\it et~al} wrote several papers \cite{Hata:2007mb,Hashimoto:2008zw} about
properties of such holographic baryons, and eventually 
\cite{Hashimoto:2009ys} worked out a repulsive force between two such baryons.
But they could not get an attractive force because of the accidental ${\bf Z}_2$
symmetry of the antipodal configuration of the flavor branes.
To see the connection, note that in the large $N_c$ limit, the nuclear forces
are dominated by the single-meson-exchange diagrams; the repulsive central forces
come from exchanges of the vector mesons while the attractive central forces
come from the scalar mesons.
In holography, the 4D vector mesons are modes of the gauge fields living on the
flavor branes, while the 4D scalar fields are modes of the scalar fields
parametrizing transverse motion of those branes.
In the original version of the Sakai--Sugimoto model, the D8 and $\overline{\rm D8}$
branes cross the color D4 branes at antipodal points of the $S^1$ circle,
hence the name ``antipodal model''.
On the gravity side of the gauge-gravity duality, the $S^1$ becomes the circular
dimension of the cigar, and the combined $\rm D8+\overline{D8}$ branes stretch
along the cigar's diameter.
The two sides of this diameter are symmetric, and this leads to the $\Phi\to-\Phi$
symmetry of the 9D scalar fields parametrizing the transverse motion of
the flavor branes;
in 4D terms, this ${\bf Z}_2$ symmetry flips the signs of all the scalar meson fields,
$\phi_i(x)\to -\phi_i(x)$.
But this symmetry does not affect the  gauge fields on the flavor branes and hence
the 4D vector mesons or the holographic baryons.
Consequently, in the antipodal model, the baryons have Yukawa couplings to
the vector mesons but not to the scalar mesons, and that's why there are no
attractive nuclear forces but only the repulsive forces.

In this article, we investigate nuclear forces in the non-antipodal version
\cite{Aharony:2006da} of the Sakai--Sugimoto model.
Without the accidental ${\bf Z}_2$ symmetry, the baryons should have Yukawa couplings
to both  vector and scalar mesons, and indeed we find both repulsive and attractive
forces.
Unfortunately, the attractive forces are too weak and the net force is repulsive at
all distances, so our model of HQCD is not too realistic.
Specifically, at intermediate distances $r$ between two nucleons ---
shorter than the ranges of the 4D Yukawa forces but longer than the size of a
baryon's core\footnote{%
	In all versions of the Sakai--Sugimoto model, the 5D instanton
	at the core of a baryon has a very small size $\rho\sim R_{KK}/\sqrt\lambda$
	where $R_{KK}$ is the Kaluza--Klein scale of extra dimensions and
	$\lambda=N_c g_{\rm YM}^2\gg1$ is the 't~Hooft coupling.
	On the other hand, the 4D mesons have masses $M_{\rm meson}\sim1/R_{KK}$
	so the Yukawa forces have ranges $R_{\rm Yukawa}\sim R_{KK}\gg\rho$.
	This is quite different from the real-life mesons and baryons where
	$\rho_{\rm baryon}\sim 4R_{\rm Yukawa}$.
	}
--- both the repulsive and the attractive potentials behave like 5D
Coulomb potential and scale like $1/r^2$.
But the attractive potential has a smaller coefficient,
\be
\rar\ \equiv\ {-V^{\rm attractive}\over V^{\rm repulsive}}\
=\ \frac19\times\left(1-\zeta^{-3}\right),\qquad
\label{Vratio}
\ee
where $\zeta\ge1$ parametrizes the geometry of the flavor D8 branes:
The near-antipodal models have $\zeta\approx1$ and $\rar\ll1$
while the far-from-antipodal models have $\zeta\gg1$ and $\rar\approx\frac19$.
In any case, $\rar<\frac19<1$ and the attractive nuclear potential is weaker
than the repulsive.

At longer distances, nuclear forces are dominated
by the 4D Yukawa potentials of the lightest mesons with the right quantum numbers,
$J^{PC}=1^{--}$ for the repulsive force and $J^{PC}=0^{++}$ for the
attractive force, thus
\be
V^{\rm repulsive}\ \propto\ +{\exp(-r\times m^{\rm lightest}_{\rm vector})\over r}\,,\qquad
V^{\rm attractive}\ \propto\ -{\exp(-r\times m^{\rm lightest}_{\rm scalar})\over r}\,.
\label{Ranges}
\ee
In real life, the lightest isoscalar scalar meson $\sigma(600)$ is lighter than the
lightest isoscalar vector meson $\omega(787)$, so at long distances the attraction
wins over the repulsion.\footnote{%
	Actually, since in real life $N_c=3\ll\infty$, the longest-range component
	of the attractive force is not the single-sigma-meson exchange but rather
	the double-pion exchange.
	The Yukawa range of this force is $1/2m_\pi$, which is significantly longer
	than $1/m_\sigma$.
	}
But in the Sakai--Sugimoto models --- antipodal or non-antipodal --- the lightest
scalar meson has more than twice the mass of the lightest vector.
Consequently, the attractive force has a shorter range than the repulsive force,
and the net nuclear force is repulsive at all distances.

We don't know why the meson spectra --- and hence the nuclear forces ---
in the Sakai--Sugimoto model are so unrealistic.
It could be something peculiar to the model's setup, hopefully to be remedied by
some future holographic models.
But it could also be a general problem of the large $N_c$ limit;
indeed, the QCD origin of the $\sigma(600)$ meson is poorly understood, and it's
not clear if for $N_c\to\infty$ it continues to exist or disappears from the spectrum.
The best way to resolve this issue
would be to find the $\sigma$ resonance and its mass
in a lattice QCD calculation for several values of $N_c$, then
extrapolate to $N_c\to\infty$.
Alternatively, once we have several different holographic models,
we can compare their predictions for the meson spectra in general
and for the lightest scalar meson in particular.
Either way, this issue will have to wait for future research.

The rest of this paper is organized as follows.
In the next section (\S2) we explain the problems with the large $N_c$
limit of nuclear physics.
First, we explain why large $N_c$ makes the nuclear matter solid rather
than liquid.
Next, we discuss the $N_c\to\infty$ limit of the nuclear forces
and what happens to the $\sigma(600)$ meson.
Finally, we bring up the issue of separating nucleons from other
baryonic species such as $\Delta$.

Section 3 is a review of the Sakai--Sugimoto model and its antipodal
and non-antipodal versions.
In particular, we derive the effective 5D Lagrangian for the $U(2)$
flavor gauge fields (for simplicity we work with two flavors), then
realize a holographic baryon as a lowest-energy YM instanton and calculate
its mass and radius.
Section 4 explains general properties of the holographic
nuclear forces in the near, intermediate, and far zones;
the three zones are illustrated in the diagram~(\ref{zones})
on page~\pageref{zones}.
We also summarize the calculation by Hashimoto {\it et~al} \cite{Hashimoto:2009ys}
of the repulsive force in the intermediate zone.

Section 5 is the core of this paper, that's where we calculate
the attractive and repulsive nuclear forces in the intermediate and far zones. 
In \S5.1 we derive the effective 5D theory of scalar and vector fields
living on the flavor branes.
We show that the abelian vector and scalar fields give rise to 5D Coulomb
forces between $SU(2)$ instantons.
In the near and intermediate zones, both the repulsive potential due to
abelian vector and the attractive potential due to the abelian scalar have
the same $1/r^2$ dependence, but the attractive potential has a smaller
coefficient as in eq.~(\ref{Vratio}).
In \S5.2 we leverage this result to obtain both isoscalar and isovector
forces between two spinning nucleons at intermediate distances from each other.
The isovector spin-spin and tensor forces stem from the small overlap
between the $SU(2)$ instantons implementing the two nucleons and their interactions
with the abelian vector and scalar fields.
Our analysis follows Hashimoto {\it et~al} \cite{Hashimoto:2009ys}, but taking the
scalar fields into account reduces the isovector forces by the same
overall factor $(1-\rar)$ as the net isoscalar $\rm repulsive-attractive$ force.
Thus,
\be
V_{\rm net}(r,I_1,I_2,J_1,J_2)\
=\ {\red(1-\rar)}\times V_{\rm net}(r,I_1,I_2,J_1,J_2)[{\rm\blue without\ scalar\ fields}].
\ee
In \S5.3 we consider the attractive forces in the far zone.
Since in the Sakai--Sugimoto model the lightest scalar meson is heavier
than the lightest vector meson, the attractive force decays with distance $r$
faster than the repulsive force, so the net isoscalar central force
is always repulsive.
We also consider the long-range isovector tensor force due to pions.
Although the pions are zero modes of the 5D vector fields and have
nothing to do with the 5D scalar fields, the pion-nucleon coupling
depends on the baryon's radius $\rho$ which is affected by the scalar-mediated
forces in the near zone.
Consequently, the isovector force due to pion exchange is reduced
by the overall factor $(1-\rar$).
Likewise, all other isovector forces in the far and intermediate zones
are reduced by the same overall factor.

Our calculation in \S5 are based on Yang--Mills approximation for the
effective Lagrangian for the flavor gauge fields.
In section~6 we investigate the validity of this approximation
by working with a complete non-abelian $\rm DBI+CS$ Lagrangian
for the fields on the flavor branes.
We show that although the $SU(2)$ gauge fields become strong
($2\pi\alpha'\CF_{MN}\sim g_{MN}$) near the center of a baryon,
the self-duality of those fields (in four non-compact space dimensions
of the D8 branes) leads to cancellation of all the higher-order
$\tr(\CF^4)$, {\it etc.,} terms in the expansion of the DBI Lagrangian.
Consequently, all our calculation in \S5 are valid to the leading order
in $1/\lambda$.
The leading  post--YM effect in 5D is a small ($O(1/\lambda)$) correction
to the self-duality condition for the $SU(2)$ gauge fields due to abelian
vector and scalar fields.
To see how this correction affects a stand-alone semiclassical baryon, in \S6
we minimize the $\rm DBI+CS$ action of an $SO(4)$--symmetric instanton-like
field configuration with a general radial profile and show that the
minimum is very close to good old YM instanton of the same radius as we
had in \S5.
Calculating  the non-abelian DBI action involves computing a symmetrized
trace; this is done in the Appendix.

Finally, section 7 summarizes our results and makes suggestions for future research.

%
\section{Limitations of the $N_c\to\infty$ Limit and Holography}
The large $N_c$ limit is inherent in all holographic QCD methods,
and this poses a problem for the aspects of nuclear physics that
are different between the small $N_c$ and the large $N_c$ regimes.
In particular, the bulk nuclear matter at zero temperature and pressure
(but finite density)
forms a quantum liquid for small $N_c$ --- such as real-life $N_c=3$
--- but becomes a crystalline solid for large $N_c$.

To see how this works, consider a condensed matter analogy --- some atoms
which attract to each other at long or medium distances but have
repulsive hard cores.
Semi-classically, at zero temperature and pressure such atoms always form
some kind of a crystal; it takes strong quantum effects to put the
atoms into some other phase such as liquid or super-solid.
Of particular importance is the kinetic energy of the zero-point
quantum motion of atoms confined to narrow potential wells,
\be
K\ \sim\ {\pi^2\hbar^2\over2 M_{\rm atom}({\rm well~diameter})^2}\,,
\label{Kwell}
\ee
or rather its ratio $K/U$ to the potential binding energy $U$ per atom.
According to Newton Bernardes \cite{Bernardes:1960}, this ratio is related
to the de~Bour parameter $\Lambda_B$ of the inter-atomic potential as
\be
{K\over U}\ \approx\ 11\Lambda_B^2\,,\qquad
\Lambda_B\ =\ {\hbar\over r_c\sqrt{2M\epsilon}}\,,
\ee
where $r_c$ is the radius of the atomic hard core and $\epsilon$ is
the maximal depth of the potential.
For small de~Bour parameters,  the quantum corrections to
the semi-classical approximation are weak and the
crystal remains stable at zero pressure.
For larger $\Lambda_B$, the  quantum corrections due to kinetic energy become
important, and when $\Lambda_B$ exceeds a critical value somewhere between 0.2
and 0.3 \cite{Glyde:1994},  the crystal melts into a quantum liquid.\footnote{%
	Melting releases the individual atoms form narrow potential wells,
	which significantly lowers their kinetic energies~(\ref{Kwell}).
	It also, moves the atoms away from the minima of the attractive potential,
	which lowers the potential binding energy $U$.
	The overall effect on the net $K-U$ energy per atom depends on the
	$K/U$ ratio: For low ratios the potential energy is more important
	and the crystal is stable, but for high ratios lowering the kinetic
	energy becomes advantageous and the crystal melts.%
	}
For example, helium atoms have $\Lambda_B=0.306$ and hence $K/U\approx1$
while neon atoms have $\Lambda_B=0.063$ and hence $K/U\approx 0.05$;
consequently, at zero temperature and zero pressure helium is a quantum liquid while
neon is a crystalline solid.

To see how the $K/U$ ratio of the nuclear matter depends on the number of colors,
we note that in the large $N_c$ limit,
the leading nuclear forces are proportional to $N_c$.
Specifically, according to Kaplan and Manohar \cite{Kaplan:1996rk},
\begin{align}
V(\vec r,I_1,I_2,J_2,J_2;N_c)\ &
=\ N_c\times A_C(r)
+\ N_c\times A_S(r)({\bf I}_1{\bf I}_2)({\bf J}_1{\bf J}_2)\nonumber\\
&\qquad+\ N_c\times A_T(r)({\bf I}_1{\bf I}_2)
		\bigl[3({\bf nJ}_1)({\bf nJ}_2)-({\bf J}_1{\bf J}_2)\bigr]
\label{NuclearPotential}\\
&\qquad+\ O(1/N_c).\nonumber
\end{align}
for the same $N_c$-independent radial profiles $A_C$, $A_S$, $A_T$ of the
central, spin-spin, and tensor potentials.
Classically, such potentials would like to arrange a many-nucleon system in
some kind of a crystal with $N_c$--independent nearest-neighbor distance${}~\sim 1$~fm,
while the binding energy of a nucleon in such a crystal would
be proportional to the $N_c$.
Indeed, all models of nuclear matter based on semi-classical models of nucleons
form such crystals, for example skyrmion crystals of ref.~\cite{Klebanov:1985qi}.
In the quantum theory, nucleons in such a lattice have zero-point
kinetic energies (\ref{Kwell}) where the well diameter is independent on $N_c$
while the nucleon's mass $M\propto N_c$, hence $K\propto1/N_c$ and
\be
{K\over U}\ \propto\ {N_c^{-1}\over N_c^{+1}}\ =\ {1\over N_c^2}\,.
\ee
We may estimate the coefficient of this proportionality using the de~Bour
parameter~$\Lambda_B$.
The maximal depth of the central potential between two nucleons is about
100~MeV for $N_c=3$, so we take it to be $\epsilon\sim N_c\times30$~MeV for large $N_c$.
Likewise, we take the nucleon mass to be $M_N\sim N_c\times 300$~MeV and
hard-core radius $r_c\sim 0.7$~fm regardless of $N_c$.
Consequently,
\be
\Lambda_B\ =\ {\hbar\over r_c\sqrt{2M\epsilon}}\
\sim\ {2\over N_c}\quad\Longrightarrow\quad
{K\over U}\sim\ {45\over N_c^2}
\label{KVratio}
\ee
and hence liquid nuclear matter for $N_c\lesssim 8$ and solid nuclear matter
for $N_c\gtrsim8$.

The numerical coefficient in eq.~(\ref{KVratio}) and hence
our estimate $N_c^{\rm crit}\sim8$ for the dividing line between liquid and solid
bulk nuclear matter (at low pressures and temperatures) should be taken with
a large grain of salt.
Also, the transition between liquid nuclear matter for $N_c=3$ and crystalline
nuclear matter for large $N_c$ may go through some exotic phases at
intermediate values of $N_c$, perhaps something like a quantum supersolid, perhaps
something more exotic without known condensed-matter analogues.
But regardless of the details of this transition, in the large $N_c$ limit
the potential energy of interacting near-static nucleons becomes much
larger than the nucleons' kinetic energies, and the bulk nuclear matter
at $T=0,P=0$ conditions becomes a conventional semi-classical crystal.
The structure of such crystals can be modeled holographically ---
and indeed there is active research in this direction (for instance \cite{Rho:2009ym})
--- but we have no experimental data to compare to the models because
real-life nuclei with $N_c=3$ are liquid rather than solid.

Meanwhile, instead of trying do build holographic models of complete
nuclei we focus on holographic models of the nuclear forces.
But even at the level of the two-body forces, the large $N_c$ limit
maybe different from the real-life case of just 3 colors.
Of particular concern is the isoscalar attractive force due to exchanges
of the $\sigma(600)$ scalar mesons between the nucleons.
In real life, this is a major component of the net attractive force
--- especially at the medium-long distances between the nucleons ---
but in the large $N_c$ limit this component may weaken or disappear because
the $\sigma(600)$ meson itself may become heavier or even disappear from
the scalar meson spectrum.

The $\sigma(600)$ (also known as $f_0(600)$) is the lightest 
isoscalar true-scalar meson.
In real life, it appears as a very broad resonance of two pions --- so broad
that its central mass is somewhat controversial and
different experimentalists locate it anywhere between 400~meV and 700~MeV,
and sometimes even higher,
{\it cf.}\ references in the Particle Data Group's listing \cite{PDGf0}.
But the real controversy about  the $\sigma(600)$ resonance is its
physical origin.
Unlike the heavier $I^G=0^+$, $J^{CP}=0^{++}$ mesons $f_0(980)$, $f_0(1370)$, {\it etc.,}
the $\sigma(600)$ meson does not exist in the non-relativistic quark model\footnote{%
	In the non-relativistic quark model, all $0^{++}$ mesons have
	$S=1$ and $L=1$.
	Consequently, the lightest $0^{++}$ meson should be heavier than
	the lightest $1^{--}$ mesons $\rho(770)$ or $\omega(787)$
	that have $S=1$ but $L=0$.
	Depending on the assumptions one makes about the forces between the
	quark and the antiquark, this argument identifies the lightest true $q\bar q$
	meson with $I^G=0^+$ and $J^{PC}=0^{++}$ as either $f_0(980)$ or $f_0(1370)$.
	In any case, the $\sigma(600)$ resonance is way too light to be a p-wave
	$q\bar q$ state, so it has to be something else.
	}
so for many years R.~L.\ Jaffe and others
\cite{Jaffe:1976fw,Jaffe:2004ph,Achasov:2007fz,Hooft:2008we} were claiming that
the $\sigma(600)$ is not a true $q\bar q$ meson but a $qq\bar q\bar q$ tetraquark.
Specifically, it's a  molecule-like bound state of two pions which exists
because the $\rho$--meson exchanges in the $t$-channel induce an attractive
$s$-channel force between the pions.
If this claim is true, then  the $\sigma$ resonance
goes away in the large $N_c$ limit because the forces between pions become
weak as $1/N_c$.

But many other authors (see \cite{KEKconference} for a sample)
identify the $\sigma(600)$ with the $\sigma$ field of the
linear sigma model of the chiral symmetry breaking.
Or rather, the massive $\sigma(x)$ field parametrizing fluctuations
of magnitude of the symmetry-breaking VEV $\langle{\overline\Psi\Psi}\rangle$
gives rise to primordial sigma quanta,
while the real sigma mesons $\sigma(600)$ are quantum mixtures
of those primordial quanta with the $\ket{\pi\pi}$ states
(and to lesser extent with the	other $I^G=0^+$, $J^{PC}=0^{++}$ mesons).
{}From this point of view, the non-relativistic quark model is irrelevant
because the quarks do not become non-relativistic until after
the chiral symmetry has already been broken.
Indeed, the NRQM does not see that the pions are (pseudo) Goldstone bosons,
so the fact that it does not see the sigma meson at all is simply another
limitation of the NRQM as far as the chiral symmetry breaking is concerned.
If this point of view is right, then  the sigma meson exits for all $N_c$.
For large $N_c$ limit, this meson is mostly a quantum of the $\sigma(x)$ field
--- its mixing with $\ket{\pi\pi}$ and other states becomes weak ---
and it's a narrow resonance rather than a broad hump we have for $N_c=3$,
but it remains a dominant resonance in the $I^G=0^+$, $J^{PC}=0^{++}$ $\pi\pi$
channel, and its mass should not be too different from the real-life 600~MeV.

The other mesons --- scalar or vector, isoscalar or isovector --- are unlikely to be disturbed
by the large $N_c$ limit, so their contributions to the nuclear forces
would be similar to real-life QCD.
{\bf IF} the $\sigma(600)$ meson remains in the spectrum in the large $N_c$
limit {\bf and if} its mass remains similar to the real-life 600~MeV,
{\bf then} the entire nuclear potential (\ref{NuclearPotential})
for $N_c\to\infty$ would be similar to what it is in real life,
except for the overall factor $N_c$.
In particular, the net central potential $V_c(r)$ would be repulsive
at short distances (the hard core) but attractive at medium and long distances:
\be
\psset{unit=7.5mm,plotpoints=300,linewidth=0.7pt}
\begin{pspicture}[shift=-5](0,-2)(12.5,+6.5)
\psline{->}(0,0)(12,0)
\rput[lb](12.1,0){$r$}
\psline{->}(0,-2)(0,+6)
\rput[lb](0,6.1){$V_c(r)$}
\psplot[linecolor=blue,linewidth=1.5pt]{1.1}{12}%
	{x -1 mul 2.71828 exch exp 1.5 mul x -0.75 mul 2.71828 exch exp sub 100 mul x div %
	 x -0.25 mul 2.71828 exch exp x mul 0.2 mul sub}
\rput[l](4,3){$N_c\to\infty$ limit with a light $\sigma$ meson}
\end{pspicture}
\label{RealForce}
\ee
On the other hand, {\bf if} the $\sigma(600)$ meson disappears from the spectrum
for large $N_c$, {\bf or if} it becomes heavier than the lightest vector meson,
{\bf then} the dominant attractive force would become shorter-ranged than the
repulsive force, and the net force at medium and long distances would be repulsive
rather than attractive:
\be
\psset{unit=7.5mm,plotpoints=300,linewidth=0.7pt}
\begin{pspicture}[shift=-5](0,-2)(12.5,+6.5)
\psline{->}(0,0)(12,0)
\rput[lb](12.1,0){$r$}
\psline{->}(0,-2)(0,+6)
\rput[lb](0,6.1){$V_c(r)$}
\psplot[linecolor=red,linewidth=1.5pt]{2}{12}%
	{x -1 mul 2.71828 exch exp 1.5 mul x -1.25 mul 2.71828 exch exp sub 100 mul x div}
\rput[l](4,3){$N_c\to\infty$ limit without the $\sigma$ meson}
\end{pspicture}
\label{BadForce}
\ee
In this scenario, at large $N_c$ the nuclear force is repulsive at all distances,
and there are no bound nuclei at all, liquid or crystalline.

So what really happens to the sigma-meson and to the nuclear forces at large $N_c$?
The best way to settle this controversy
would be to find the $\sigma$ resonance and its mass
in a lattice QCD calculation for several values of $N_c$.
Such a calculation would  require a realistic pion mass
(unlike most present-day lattice calculations extrapolating from $m_\pi\ge 350$~MeV)
and rather large lattices to distinguish the sigma resonance from the two-pion
continuum, so it may be too hard for the present-day computers.
But thanks to the Moore's Law, finding the $\sigma$ resonance on a lattice
should become possible in a not-too-distant future.

Alternatively, we may try to resolve the issue using holography.
Although a holographic model of real QCD --- or rather, of QCD with large~$N_c$---
is yet to be constructed, several known models seem to be qualitatively similar,
so we can compare their predictions for the meson spectra in general,
and for the lightest true scalar meson in particular.
However, the models that {\it seem qualitatively similar} to QCD may not be similar
enough, and their predictions could be widely off target.
Indeed, the predictions of different models have turned out to be quite different from each other.
For example, in the Sakai--Sugimoto model which we use in this article,
the lightest true scalar meson is more than twice as heavy as the lightest vector meson.
Consequently --- as we shall see in painful detail in section~5 ---
the net nuclear force is everywhere repulsive and looks like~(\ref{BadForce})
rather than like~(\ref{RealForce}).
On the other hand, in the highly-non-antipodal version of the
Dymarsky--Kuperstein--Sonnenschein model \cite{Banerjee:2001js},
the lightest $J^{CP}=0^{++}$ meson is much lighter than any other mesons
(except the pions) \cite{Dymarsky:2010ci}.
However, this lightest scalar is a pseudo-Goldstone boson of the
approximate conformal symmetry of the flavor sector, so it is not clear how
much attractive force it can mediate.
As of this writing, it is not clear if the net nuclear potential in this
model looks like the real-life potential~(\ref{RealForce}) or like the
everywhere-repulsive potential~(\ref{BadForce}) 
we calculate in this paper for the Sakai--Sugimoto model.

But suppose tomorrow somebody discovers a holographic model of the real QCD
and --- miracle of miracles --- it has a realistic spectrum of mesons, including the
$\sigma(600)$ resonance, and even the realistic Yukawa couplings of those mesons
to the baryons.
Even for such a model, the two-body nuclear forces would not be quite as in
the real world because the  semi-classical holography limits
$N_c\to\infty$, $\lambda\to\infty$ suppress the multiple meson exchanges
between baryons.
Although in this case, the culprit is not the large number of colors but the
large 't~Hooft coupling $\lambda=N_c g^2_{\rm YM}$.

Indeed, from the hadronic point of view, nuclear forces
arise from the mucleons exchanging one, two, or more mesons,
and in real life the double-meson exchanges are just as important
as the single-meson exchanges.
In particular, since the lightest mesonic state with $I^G=0^+$, $J^{CP}=0^{++}$
quantum numbers is a pair of un-bound pions, the longest-range isoscalar attractive
force between nucleons comes from exchanges of two pions rather than of any
single mesons.
In holography, the single-meson exchanges happen at the tree level of the string theory
while the multiple meson exchanges involve string loops ($k-1$ loops for $k$ mesons),
and  the loop amplitudes are suppressed by the powers of $1/\lambda$
relative to the tree amplitudes.

Naively, one would expect the loop amplitudes to carry additional factors of $1/N_c$
(which is dual to the string coupling) rather than $1/\lambda$, or maybe both
$1/\lambda$ and $1/N_c$ factors, but the naive
power-of--$N_c$ counting does not work for loop amplitudes
involving baryons made of $N_c$ quarks.\footnote{%
    The authors thank Alexei Cherman and Thomas Cohen for bringing this
    fact to our attention after we made a mistake in the first version of this paper.
    }
Indeed, in honest QCD with a large number of colors, the multi-meson-exchange
contributions to the non-relativistic effective potential for the baryons
are {\bf not} suppressed by powers of $1/N_c$ \cite{Banerjee:2001js}.
However, the extra powers of $N_c$ due to $N_c$ quarks in a baryon are not
accompanied by the extra powers of $\lambda$, so in holography,
the contributions of the multiple meson exchanges are suppressed,
albeit by powers of $1/\lambda$ rather than $1/N_c$.

To see how this works in a general holographic model of QCD
with $N_c\gg N_f$, note that such a model starts with a string-theoretic
construction where the colors and the flavors live on separate branes.
For large $N_c$ and large $\lambda$, the color branes become black
branes producing curvature and fluxes through the bulk, which provide
a non-trivial background for degrees of freedom living in the bulk itself
as well as on the flavor branes.
The bulk degrees of freedom are dual to the pure-color sector of QCD (glueballs, {\it etc.}),
while the vector and scalar fields living on the flavor branes are dual
to the $q\bar q$  mesons.
The flavor fields have rather weak couplings to each other:
in 5D terms,
\be
g^{}_{\rm 5D,flavor}\ \sim\ {\sqrt{R_{KK}}\over\sqrt{\lambda N_c}}
\ee
so the 4D mesons --- which are modes of the 5D vector and scalar fields
with wave functions $\psi\sim R_{KK}^{-1/2}$ --- have couplings to each other
of the order
\be
\begin{pspicture}[shift=-1.73](-2,-1.8)(+1,+1.8)
\psline[linestyle=dashed,linecolor=blue](-2,0)(0,0)
\psline[linestyle=dashed,linecolor=blue](+1,+1.73)(0,0)(+1,-1.73)
\pscircle*(0,0){0.1}
\end{pspicture}\quad
g_{MMM}^{}\ \sim\ {1\over\sqrt{\lambda N_c}}\,.
\label{3mesons}
\ee
A holographic baryon is made from some brane spanning only the compact dimensions that
is connected to the flavor branes by $N_c$ strings, although this construction is often
equivalent to an instanton of the 5D flavor gauge fields.
Consequently, the baryon-meson coupling is enhanced by an extra factor of $N_c$,
\be
\psset{unit=9mm,arrowscale=1}
\begin{pspicture}[shift=-1.6](-2,-1.7)(+1,+1.7)
\psline[linestyle=dashed,linecolor=blue](-2,0)(0,0)
\psline[doubleline=true,linecolor=red]{>->}(+1,+1.73)(0,0)(+1,-1.73)
\pscircle*(0,0){0.1}
\end{pspicture}\quad
g_{MB\bar B}^{}\ \sim\ N_c\times {1\over\sqrt{\lambda N_c}}\
=\ {\sqrt{N_c}\over\sqrt{\lambda}}\,.
\label{MBB}
\ee

At the tree level of the baryon-meson theory, scattering of two baryons proceeds
through a single-meson exchange, which produces a $O(N_c/\lambda)$ amplitude,
\be
\psset{unit=9mm,arrowscale=1}
\begin{pspicture}[shift=-1.6](-3,-1.7)(+3,+1.7)
\psline[linestyle=dashed,linecolor=blue](-1,0)(+1,0)
\psline[doubleline=true,linecolor=red]{>->}(+2,+1.73)(+1,0)(+2,-1.73)
\psline[doubleline=true,linecolor=red]{>->}(-2,+1.73)(-1,0)(-2,-1.73)
\pscircle*(+1,0){0.1}
\pscircle*(-1,0){0.1}
\end{pspicture}\quad
{\cal A}^{\rm tree}\sim\ g^2_{MB\bar B}\ \sim\ {N_c\over\lambda}\,.
\label{SingleExchange}
\ee
At the one-loop level, there are two types of diagrams, the triangle diagrams such as
\be
\psset{unit=9mm,arrowscale=1}
\begin{pspicture}[shift=-2.4](-3,-2.5)(+4,+2.5)
\psline[doubleline=true,linecolor=red]{>->}(-2,+2.5)(-1,0)(-2,-2.5)
\psline[doubleline=true,linecolor=red]{>->}(+3,+2.5)(+2,+1)(+2,-1)(+3,-2.5)
\psline[linestyle=dashed,linecolor=blue](-1,0)(+1,0)
\psline[linestyle=dashed,linecolor=blue](+2,+1)(+1,0)(+2,-1)
\pscircle*(+1,0){0.1}
\pscircle*(-1,0){0.1}
\pscircle*(+2,+1){0.1}
\pscircle*(+2,-1){0.1}
\end{pspicture}\quad
{\cal A}^\Delta\sim g^3_{MB\bar B}\times g^{}_{MMM}\ \sim\ {N_c\over\lambda^2}
\label{Triangle}
\ee
and the box and crossed-box diagrams
\be
\psset{unit=9mm,arrowscale=1}
\begin{pspicture}[shift=-2.4](-6.5,-2.5)(+6.5,+2.5)
\rput(-3.5,0){
    \psline[doubleline=true,linecolor=red]{>->}(+2,+2.5)(+1,+1)(+1,-1)(+2,-2.5)
    \psline[doubleline=true,linecolor=red]{>->}(-2,+2.5)(-1,+1)(-1,-1)(-2,-2.5)
    \psline[linestyle=dashed,linecolor=blue](-1,+1)(+1,+1)
    \psline[linestyle=dashed,linecolor=blue](-1,-1)(+1,-1)
    \pscircle*(+1,+1){0.1}
    \pscircle*(+1,-1){0.1}
    \pscircle*(-1,+1){0.1}
    \pscircle*(-1,-1){0.1}
    }
\rput(+3.5,0){
    \psline[doubleline=true,linecolor=red]{>->}(+2,+2.5)(+1,+1)(+1,-1)(+2,-2.5)
    \psline[doubleline=true,linecolor=red]{>->}(-2,+2.5)(-1,+1)(-1,-1)(-2,-2.5)
    \psline[linestyle=dashed,linecolor=blue](-1,+1)(+1,-1)
    \psline[linestyle=dashed,linecolor=blue](-1,-1)(+1,+1)
    \pscircle*(+1,+1){0.1}
    \pscircle*(+1,-1){0.1}
    \pscircle*(-1,+1){0.1}
    \pscircle*(-1,-1){0.1}
    }
\rput(0,0){
    \psline[linewidth=1pt](0,-0.3)(0,+0.3)
    \psline[linewidth=1pt](-0.3,0)(+0.3,0)
    }
\end{pspicture}
\label{BoxDiagrams}
\ee
with amplitudes
\be
{\cal A}^{\square}\ \sim\ g^4_{MB\bar B}\ \sim\ {N_c^2\over\lambda^2}\,.
\label{Abox}
\ee
that carry an extra power of $N_c$.
However, Banerjee {\it et~al}  showed \cite{Banerjee:2001js}
that for non-relativistic baryons, the box and the crossed-box diagrams
almost cancel each other from the effective potential between the baryons,
with the un-canceled part having a lower power of the $N_c$.
Banerjee {\it et~al} did not pay any attention to the powers of $\lambda$,
but clearly the un-canceled sub-leading terms in the box and crossed-box
diagrams cannot carry higher powers of the 't~Hooft coupling than the
leading terms~(\ref{Abox}), thus
\be
{\cal A}^{\square}_{\rm uncanceled}\ \sim\ {N_c\over\lambda^2}\ 
\sim\ {\cal A}^\Delta\
\sim\ {1\over\lambda}\,{\cal A}^{\rm tree}.
\ee
In other words, the contribution of the double-meson exchange and other one-loop
processes to the 2--body nuclear potential carries the same power of $N_c$
but is suppressed by a factor $1/\lambda$ compared to the tree-level
singe-meson exchange.

To be precise, the large $\lambda$ limit suppresses exchanges of the {\it un-bound}
meson pairs but not of the meson-meson resonances --- which become narrow
(because of weak $g_{MMM}^{}$) and
act as single mesons exchanged between the two baryons.
In particular, in the isoscalar $0^{++}$ channel that gives rise to the dominant
attractive force between nucleons, the $\lambda\to\infty$ limit suppresses the
contribution of the unbound two-pion continuum,
but it replaces it with a discrete set of $f_0$ resonances.
In a good holographic model of QCD (which alas has not been found yet),
{\it the overall strength} of the $0^{++}$ channel should be similar to
the real QCD, so it would produce a similar isoscalar attractive force
{\it at short distances.}
However, the range of this attractive force would be significantly shorter:
Instead of decaying with distance like $\exp(-2m_\pi r)$ as in real life,
the holographic attractive force decays as $\exp(-m_0 r)$
where $m_0$ is the mass of the lightest isoscalar $0^{++}$ meson, presumably
$\sigma$(600~MeV).

In principle, the isoscalar $1^{--}$ channel that gives rise to the dominant
repulsive force suffers from similar corrections in the $\lambda\to\infty$ limit.
But in practice, the strongest and the longest-range contribution to this channel
comes from exchanges of a single $\omega(787)$ meson, so suppressing the
multi-meson exchanges in this channel would not make a qualitative difference.
Thus altogether, the net effect of large 't~Hooft coupling on the central nuclear potential
--- besides the overall $1/\lambda$ factor --- is the shortening of the attractive tail
at long distances:
\be
\psset{unit=7.5mm,plotpoints=300,linewidth=0.7pt}
\begin{pspicture}[shift=-5](0,-2)(12.5,+6.5)
\psline{->}(0,0)(12,0)
\rput[lb](12.1,0){$r$}
\psline{->}(0,-2)(0,+6)
\rput[lb](0,6.1){$\lambda\times V_c(r)$}
\psplot[linecolor=red,linewidth=1.5pt]{1.1}{12}%
	{x -1 mul 2.71828 exch exp 1.5 mul x -0.75 mul 2.71828 exch exp sub 100 mul x div}
\psplot[linecolor=blue,linewidth=1.5pt]{1.1}{12}%
	{x -1 mul 2.71828 exch exp 1.5 mul x -0.75 mul 2.71828 exch exp sub 100 mul x div %
	 x -0.25 mul 2.71828 exch exp x mul 0.2 mul sub}
\rput[l](3,4){\blue blue: real QCD, $\lambda\sim1$}
\rput[l](3,3){\red red: best possibility for holographic QCD, $\lambda\gg1$}
\end{pspicture}
\label{GoodForce}
\ee
\par\noindent
However, this optimistic picture presumes a holographic model of QCD that
correctly reproduces (a) the overall strength of the isoscalar $0^{++}$
and $1^{--}$ channels, and (b) the mass spectra of vector and scalar mesons,
especially the  masses of the lightest $0^{++}$ and $1^{--}$ mesons
$\sigma(600)$ and $\omega(787)$.
But thus far, no known model satisfies these requirements, not even approximately,
so the nuclear forces they produce could be much more different from the real life
than~(\ref{GoodForce}).
In particular, the nuclear force we calculate in this paper for the Sakai--Sugimoto
model turns out to be everywhere repulsive:
\be
\psset{unit=7.5mm,plotpoints=300,linewidth=0.7pt}
\begin{pspicture}[shift=-5](0,-2)(12.5,+6.5)
\psline{->}(0,0)(12,0)
\rput[lb](12.1,0){$r$}
\psline{->}(0,-2)(0,+6)
\rput[lb](0,6.1){$V_c(r)/N_c$}
\psplot[linecolor=red,linewidth=1.5pt]{2}{12}%
	{x -1 mul 2.71828 exch exp 1.5 mul x -1.25 mul 2.71828 exch exp sub 100 mul x div}
\psplot[linecolor=blue,linewidth=1.5pt]{1.1}{12}%
	{x -1 mul 2.71828 exch exp 1.5 mul x -0.75 mul 2.71828 exch exp sub 100 mul x div %
	 x -0.25 mul 2.71828 exch exp x mul 0.2 mul sub}
\rput[l](4,4){\blue blue: real QCD}
\rput[l](4,3){\red red: Sakai--Sugimoto model}
\end{pspicture}
\label{SSForce}
\ee

Now let's go back to the large $N_c$ limit --- in holography or in honest QCD
--- and consider yet another general problem with baryons made from many quarks:
How to separate the nucleons with $I=J=\tfrac12$ from the other kinds of baryons
such as $\Delta$ with $I=J=\tfrac32$?
In real life, there is a large mass gap between the nucleons and the $\Delta$ baryons
--- almost 300~MeV --- but for large $N_c$ this gap shrinks as $1/N_c$.
At the same time, the two-baryon potential grows like $N_c$, so for large $N_c$
it becomes stronger  than the gap.
Consequently, two interacting nucleons may ``forget'' their individual spins
and isospins and mix up with other baryonic species such as $\Delta$.
In fact, for large $N_c$ there is a whole lot of baryonic species with
$I=J$ ranging from $\tfrac12$ (for odd $N_c)$ or 0 (for even $N_c$) all the way
up to $N_c/2$, and a strongly-interacting nucleon might mix up with all of them.
While such mixing would not affect the isoscalar spin-blind central force
between two baryons, it might significantly enhance the isovector spin-spin and
tensor forces.

Therefore, comparing the two-baryon forces in the large $N_c$ limit to the
real-life two-nucleon forces is rather tricky.
One has to carefully keep track of the spin and isospin degrees of freedom
of the two baryons,
expand the interaction Hamiltonian into central, spin-spin, and tensor forces
as in eq.~(\ref{NuclearPotential}), and then compare the radial profiles
$A_C(r)$, $A_S(r)$, and $A_T(r)$.
Moreover, the spin and isospin degrees of freedom require quantum mechanical
treatment because semi-classically,
we do not get definite spins or isospins even for stand-alone
single baryons.
Instead, we get skyrmions, or instantons, or some other kind of solitons
with a definite orientation of the $SU(2)_{\rm isospin}$ relative to the
$SU(2)_{\rm spin}$; in quantum terms, they become superpositions
of baryons with all possible $I=J=\tfrac12,\tfrac32,\ldots,\infty$.
Consequently, a force between two such semiclassical baryons is not a force
between two nucleons but rather a superposition of forces between different
baryonic species.

This problem affected the first holographic calculation of the nuclear forces
by K.~Y.~Kim and I.~Zahed \cite{Kim:2008iy}.
Their baryons were semiclassical instantons in the Sakai--Sugimoto model,
so instead of definite $\ket{I,I_z,J,J_z}$ they had a definite
direction $\bf n$ in $S^3=SU(2)_{\rm isospin}\times SU(2)_{\rm spin}/SU(2)_{\rm common}$.
Consequently, Kim and Zahed\cite{Kim:2008iy} found that the force between two baryons depends
on the angle between $\bf n_1$ and $\bf n_2$ --- it was attractive for some angles and
repulsive for other --- but they could not interpret this angular dependence
in terms of the isovector spin-spin and tensor forces.
By comparison, Hashimoto, Sakai, and Sugimoto \cite{Hashimoto:2009ys} made a similar
calculation using properly quantized collective coordinates for each instanton.
Consequently, they obtained the force between two nucleons rather than some
mixed-up baryons, and they could see how this force depends on each nucleon's
$I_z$ and $J_z$.
In particular, they saw that at medium-short distances, the net force between
two nucleons is always repulsive.
Evidently, the attraction Kim and Zahed saw for some relative orientations
of semiclassical baryons happens only for high spins and isospins, but not
for nucleons with $I=J=\tfrac12$.

On the other hand, the analysis of Hashimoto {\it et~al} was limited to
the first-order perturbation theory for nucleons that are far enough from
each other to avoid the strong mixing of spins and isospins.
This approach will not work for the hard-core region at very short
distances where the interactions are much stronger than the gaps between
states of the individual baryons.
In the hard core, the semiclassical analysis of Kim and Zahed might work better
than the perturbative expansion of Hashimoto {\it et~al}, although
comparing the semi-classical large--$N_c$ results to the real-life nuclear forces
might be problematic.

To summarize, the large $N_c$ limit of nuclear physics suffers from three
major problems.
The third problem of baryon mixing is only technical, and it can be solved
--- at least for the medium and long distances between the nucleons ---
by following  Hashimoto {\it et~al} rather than Kim and Zahed.
But there are no ways around the first problem of different phase structures
of nuclear matter with $N_c=3$ and with $N_c\to\infty$.
Even at high pressures and densities, there is a difference:
For $N_c=3$, squeezing nucleons together makes them merge into a quark liquid,
while for $N_c\to\infty$ the nucleons always retain their individual identities
and a would-be quark liquid suffers from the ``chiral density wave'' instability
\cite{BLL-RSVW}.
It is possible that at some intermediate pressures and densities the $N_c=3$
nucleons form a crystal --- just like helium solidifies at high pressures ---
before merging into a quark liquid.
If this intermediate-pressure phase of real nuclear matter is ever observed in a lab,
or can be reliably shown to exist in some exotic but observable places
like inferiors of neutron stars, it would be very interesting to compare its
properties to the holographic models.
Until then, we can only speculate.

Finally, the second problem --- concerning the fate of the $\sigma(600)$ resonance
in the large $N_c$ limit and its effect on the attractive nuclear force ---
is solvable in principle, but it has not been solved yet.
In holography, this problem is aggravated by using QCD-like models
in lieu of the presently unknown holographic dual of the real QCD.
The meson spectra of such models are not quite realistic; for example, in
the Sakai--Sugimoto model (both antipodal and non-antipodal versions)
there is no $\sigma$ resonance and the lightest scalar meson has more than twice the mass
of the lightest vector meson.
Consequently, we shall see in section~5 that in this model, the attractive force is both
weaker and shorter-ranged than the repulsive force, so the net nuclear force
is always repulsive.
This could be a peculiar failing of the Sakai--Sugimoto model, or it could
be the general problem of holography or even of the large $N_c$ limit.
Hopefully, future research will resolve this issue.

%
\section{Baryons in the Non-antipodal Sakai-Sugimoto Model}
The Sakai--Sugimoto \cite{Sakai:2004cn}  model is based on placing a set of $N_f$
$D8$ and anti-D8 flavor branes
into the gravitational background of $N_c$ coincident near-extremal
$D4$ branes \cite{Witten:1998zw}.
We take the $N_c\gg N_f$ limit, so the flavor branes are treated as probes.
The color D4 branes span the Minkowski spacetime an a compact circle of radius $R$
in the direction $x_4$.
In the holographic limit, these branes become a background comprised of the
following metric, RR 4--flux, and dilaton:
\begin{align}
ds^2\ &
=\ \left( \frac{u}{R_{D4}}\right)^{3/2}
\Bigl[-dt^2+\delta_{ij}dx^idx^j+f(u)dx_4^2\Bigr]\,
+\,\left( \frac{R_{D4}}{u}\right)^{3/2}
\left[\frac{du^2}{f(u)}+u^2d\Omega_4^2\right]\nonumber\\
F_4\ &
=\ \frac{2\pi N_c}{V_4}\times\epsilon_4\,,\qquad
e^{\phi}\ =\ g_s\left( \frac{u}{R_{D4}}\right)^{3/4},
\end{align}
where
\be
R_{D4}^3\ =\ \pi g_sN_cl_s^3\,,\qquad
f(u)\ =\ 1\,-\,\left( \frac{u_{\Lambda}}{u}\right)^3,
\ee
$V_4$ is the volume of the unit sphere $\Omega_4$ and $\epsilon_4$ is its 4--volume form,
$l_s=\sqrt{\alpha'}$ is the string length, and $g_s$ is the  string coupling.

The manifold spanned by the coordinates $u$ and $x_4$ has the topology of
a cigar with tip at $u_{\rm min}=u_{\Lambda}$;
the flavor branes span a continuous line on that cigar, see figure~(\ref{Ushape}).
In order to avoid a conical singularity at the tip of the cigar,
the radius $R$ of the $x_4$ circle has to satisfy
\be
2\pi R\ =\ \frac{4\pi}{3}\left( \frac{R_{D4}^3}{u_{\Lambda}}\right)^{1/2}.
\ee
Consequently, the Kaluza--Klein scale of the model is
\be
M_{\Lambda}\ =\ \frac{1}{R}=\frac{3}{2}\frac{u_{\Lambda}^{1/2}}{R^{3/2}_{D4}}
\ee
while the confining string tension \cite{Kinar:1998vq} is
\be
\label{stringtension}
T_{\rm str}\ =\ \frac{1}{2\pi\ell_s^2}\sqrt{g_{xx}g_{tt}}|_{u=u_{\Lambda}}\
=\ \frac{1}{2\pi\ell_s^2}\left( \frac{u_{\Lambda}}{R_{D4}}\right)^{3/2}
\ee

Besides the line on the $(u,x_4)$ cigar, the D8 branes span the Minkowski
spacetime and the compact $S^4$ sphere.
All $N_f$ branes are coincident, and the action of this brane stack has
a Dirac--Born--Infeld (DBI) term and a Chern--Simons (CS) term,
\be
S_{\rm D8} = S_{\rm DBI} + S_{\rm CS} .
\ee
The  DBI term is
\be
\label{DBIaction}
S_{\rm DBI}\ =\ T_8\!\int\! d^9x\, e^{-\phi}
\str\left(\sqrt{-\det(g_{mn} + 2\pi\alpha' \CF_{mn})}\right)
\ee
where $T_8 = (2\pi)^{-8}l_s^{-9}$ is the D8-brane's tension, $g_{mn}$
is the nine-dimensional induced metric on the branes,
\be
ds_{\rm D8}^2\
=\ \biggl({u \over R}\biggr)^{3 \over 2}\eta_{\mu\nu}dx^\mu dx^\nu\
+\ \Biggl[
	\biggl({u \over R}\biggr)^{3 \over 2}f(u)\,
    +\,\biggl({R \over u}\biggr)^{3 \over 2}{1\over f(u)}\left({du\over dx_4}\right)^2
	\Biggr]dx_4^2\
+\ \biggl({R \over u}\biggr)^{3 \over 2}u^2d\Omega_4^2 ,
\ee
$\CF_{MN}$ is the $U(N_f)$ gauge field strength
on the worldvolume of the D8-branes,
and $\str$ is the symmetrized trace over the flavor indices.
The 9D Chern--Simons term involves the 5D gauge fields and the RR flux on the $S^4$;
after integration over the $S^4$ it becomes the usual 5D CS term,
\be
S_{\rm CS}\ =\ {N_c \over 24\pi^2} \int\!
\tr\left (\CA\CF^2-{i \over 2}\CA^3\CF-{1 \over 10}\CA^5\right ) .
\ee

In the geometry of the cigar, the $D8$ and the $\bar {D8}$ branes cannot continue indefinitely
in the negative $u$ directions, and they have no place to end.
Instead, the branes connect to the antibranes and form continuous U-shaped lines,
whose geometry satisfies classical equation of motion for the DBI action~(\ref{DBIaction}).
The solutions form a family parametrized by the $u_0\ge u_\Lambda$, the lowest point
on the brane stack.
Figure (\ref{Ushape}) below illustrates two such solutions:
\begin{figure}[t]
\begin{center}
\vspace{3ex}
\includegraphics[width= 100mm]{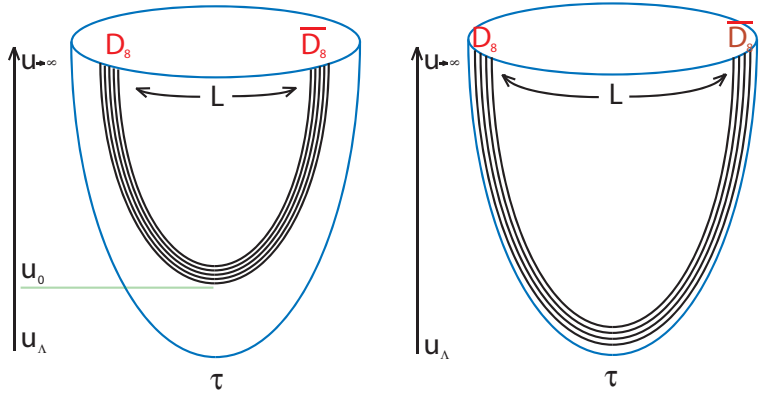}\end{center}
\caption{The figure on the right is the generalized  non-antipodal configuration.
The figure
on the right describes the limiting antipodal case $L=\pi R$, where
the branes connect at $u_0=u_{\Lambda}$.
\label{Ushape}}
\end{figure}
The physical meaning of the $u_0$ parameter is the ``string endpoint mass"
 of a quark defined such that the mass of a stringy
meson involving a non-spinning string of length $L$ is
$M_{\rm meson}= T_{\rm str}L+ 2 m_q^{\rm SEP}$.
In terms of the brane geometry, this endpoint mass is \cite{Mintakevich:2008mm}
\be
\label{constituentmass}
m_q^{\rm SEP}\ =\ \frac{1}{2\pi\alpha'}
\int^{u_0}_{u_{\Lambda}}\sqrt{-g_{tt}g_{uu}}\,du\
=\ \frac{1}{2\pi\alpha'}\int^{u_0}_{u_{\Lambda}}f^{-1/2}(u)\,du .
\ee
The original Sakai--Sugimoto model  assumed the antipodal configuration of the 
flavor branes, as shown on the right side of the figure~(\ref{Ushape});
for this configuration $u_0=u_\Lambda$ and $m^{\rm SEP}=0$.
But comparing holographic predictions to for the meson and baryon masses to the
experimental data shows that non-zero end-point mass gives a better fit.
Likewise, the width of a meson due to its decay into two mesons
\cite{Peeters:2005fq}, the  best fit to Regge trajectories,
and matching baryonic properties \cite{Seki:2008mu}, all favor non-zero end-point masses
and hence non-antipodal models depicted on the left side
of the figure~(\ref{Ushape}).

In section 6 we shall consider the full $\rm DBI+CS$ action for the flavor branes,
but most of our analysis will be based on the weak-gauge-field approximation
\be
S_{\rm DBI}[g,\CF]\ =\ S^0_{\rm DBI}[g]\ +\ S_{\rm YM}\ +\ {\cal O}(\CF^4)
\label{SDBIexpand}
\ee
where
\be
S^0_{\rm DBI}\ =\ {N_f T_8 V_4 \over g_s} \int d^4x\, dx_4\,
    \sqrt{u^8f(u) + {R_{D4}^3\,u^5 \over f(u)}\left({du\over dx_4}\right)^2}
\ee
is the action for the metric with switched-off gauge fields.
To simplify this action --- as well as the YM action for the gauge fields ---
it's convenient to change the cigar coordinate along the branes from $u$ or $x_4$
to $z$ defined according to \cite{Seki:2008mu}
\be
\label{zcoord}
u\ =\ u_{\rm \Lambda}(\zeta^3 +\zeta z^2)^{1/3} ,\qquad
\zeta\ =\ \frac{u_0}{u_{\rm \Lambda}}\,.
\ee
The reason for this choice is that both $x_4(z)$ and $u(z)$ are single-valued
smooth functions (unlike the $x_4(u)$ function which is double-valued and singular
at $u-u_0$).
Unlike in  \cite{Sakai:2004cn}, here $z$ is  dimensionless;
its values run from $-\infty$ to $+\infty$.
The $\zeta$ parameter is 1 for the antipodal model while the non-antipodal
models have $\zeta>1$.

The Yang-Mills part of the action (\ref{SDBIexpand})  is obtained by expanding
(\ref{DBIaction}) to the lowest non-trivial order in gauge fields.
In terms of the $z$ coordinate,
\be
\label{actionz}
S_{\rm YM}\ =\ -\kappa\! \int\!\!d^4x\,dz\,
\tr\Bigl(\tfrac12 h(z;\zeta)\CF_{\mu\nu}^2\,
+\,M_{\rm \Lambda}^2 k(z;\zeta)\CF_{\mu z}^2\Bigr)
\ee
where
\be
\kappa\ =\ \lambda N_c\frac{1}{(216\pi^3)}\,,\quad
\lambda\ =\ N_c g_{YM}^2\,,\quad
h(z;1)\ =\ (1+z^2)^{-1/3},\quad k(z;1)\ =\ 1+z^2,
\ee
and the full expressions for $h(z;\zeta) $ and $ k(z;\zeta)$ with $\zeta>1$
are given in equation (4.2) of \cite{Seki:2008mu}.

For the case of two flavors,  the  $U(2)$ gauge fields can be decomposed as
\be
\CA^m\ =\ A^m_{\rm SU(2)}\ +\ \tfrac12\hat A^m_{U(1)}\
=\ \tfrac12 A^{ma}\times\tau^a\ +\ \tfrac12\hat A^m\times{\bf 1}_{2\times2}
\ee
where $A^m$ are the $SU(2)$ gauge fields, $\tau^a$ are the Pauli matrices,
and $\hat A^m$ is the abelian gauge field in $U(1)\subset U(2)$.
For a baryon, the $SU(2)$ gauge fields form an instanton in the 4 space dimensions
$(1,2,3,z)$, while Chern--Simons coupling induces the abelian electric field $\hat A_0$.
After rescaling the coordinates and truncating equations of motion
to the leading terms in $1/\lambda$, the gauge fields take form
\cite{Hashimoto:2009ys,Seki:2008mu}
\begingroup
\interdisplaylinepenalty=1000
\begin{align}
\label{BaryonFields}
\begin{split}
A^{\rm cl}_M(x^i,\Tz)\ &
=\-i v(\xi)\times g\,\partial_M g^{-1}\\
v(\xi)\ &
=\ {\xi^2 \over \xi^2 + \rho^2}\,,\qquad
\xi\ =\ \sqrt{(x^i - X^i)^2 + (\Tz-\TZ)^2}\,,\\
g(x^i,\Tz)\ &
=\ {(\Tz-\TZ){\bf 1}\,-\,i(x^i-X^i)\tau_i \over \xi} \quad (i=1,2,3),\\
\HA^{\rm cl}_0\ &
=\ {27\pi\over \lambda\zeta}\,{\xi^2 +2\rho^2 \over (\xi^2 + \rho^2)^2}\,,\\
A^{\rm cl}_0\ &
=\ \HA^{\rm cl}_M\ =\ 0,
\end{split}
\end{align}
\endgroup
where $M,N=1,2,3,z$ and $\Tz= \frac{3}{8\zeta^3-5}\times\frac{z}{M_\Lambda}$.
Note that there is a critical value for $\zeta$, namely, $\zeta= (5/8)^{1/3}$,
but this critical value is unphysical --- all the Sakai--Sugimoto models
have $\zeta\ge1$ since the branes cannot go below the tip of the cigar.
Substituting the fields~(\ref{BaryonFields}) into the $\rm YM+CS$ action
and minimizing with respect to the baryon radius $\rho$ gives the baryon's
classical mass
\be
M_{\rm cl}\ =\ 8\pi^2\kappa \left( \zeta + {18\pi \over
\lambda\zeta}\sqrt{8\zeta^3-5 \over 10} \right)\quad
{\rm for}\quad \rho_{\rm cl}^2\ =\ {81\pi \over \lambda}\sqrt{2 \over 40\zeta^3-25}
\label{BMR}
\ee
Note that similar to the antipodal case, the baryon's radius scales as $\lambda^{-1/2}$,
only the numerical coefficient is different for $\zeta\neq1$.

By repeating the analysis of \cite{Hata:2007mb} for the non-antipodal models,
Seki and Sonnenschein \cite{Seki:2008mu} obtained
the mass spectrum for baryons with different $I=J$ and radial
quantum numbers, the mean-square charge radii, the isoscalar and isovector magnetic
moments, {\it etc., etc.,} as functions of $\zeta$ and $M_\Lambda$.
The found the best fit to experimental data obtain for an un-physical
$\zeta\approx0.942$, which indicates that the generalized Sakai--Sugimoto model
is not a very accurate description of real-life baryons. 
Nevertheless, in this article we shall stick with the
non-antipodal Sakai--Sugimoto models simply because its the only model
of holographic QCD we know in detail. 

%
\section{Summary of the Repulsive Force}
In real life, the nucleon has a fairly large radius compared to the ranges
of mesonic Yukawa forces (except pion's),
$R_{\rm nucleon}\sim 4/M_{\rho\,\rm meson}$.
But in the holographic nuclear physics with ${\lambda\gg1}$, we have the opposite
situation: While the meson masses are $O(M_\Lambda)$ and the 4D Yukawa forces have
$O(1/M_\Lambda)$ ranges, the baryon has a much smaller radius
$R_{\rm baryon}\sim \lambda^{-1/2}/M_\Lambda$, {\it cf.}\ eq.~(\ref{BMR}).
Thanks to this hierarchy, the nuclear forces between two baryons at
distance $r$ from each other fall into 3 distinct zones:
$$
\psset{unit=0.0565\displaywidth}
\begin{pspicture}(-0.7,-1.5)(17,8.2)
\psframe*[linecolor=red](0,0)(2,7.5)
\rput*(1,5){near\strut}
\rput*(1,4.2){zone\strut}
\psframe*[linecolor=yellow](2,0)(8,7.5)
\rput*(5,5){intermediate zone}
\psframe*[linecolor=green](8,0)(14,7.5)
\rput*(11,5){far zone}
\psline[linewidth=0.5pt](2,-0.5)(2,7.5)
\rput[t](2,-0.5){\psovalbox{$R_{\rm baryon}\sim1/M_\Lambda\sqrt{\lambda}$}}
\psline[linewidth=0.5pt](8,-0.5)(8,7.5)
\rput[t](8,-0.5){\psovalbox{$M_{\rm meson}^{-1}\sim1/M_\Lambda$}}
\psline{->}(0,0)(14.5,0)
\rput[lb](14.3,0.2){$r$}
\parametricplot[plotstyle=dots,dotstyle=|,dotsize=0.12 0,plotpoints=25]%
	{1}{25}{t 0.2 mul sqrt 2 mul -0.06}
\parametricplot[plotstyle=dots,dotstyle=|,dotsize=0.2 0,plotpoints=49]%
	{1}{49}{t sqrt 2 mul -0.1}
\psline{->}(0,0)(0,8)
\rput[l](0.2,8){$V$}
\rput{90}(0,0){%
	\parametricplot[plotstyle=dots,dotstyle=|,dotsize=0.12 0,plotpoints=25]%
		{1}{25}{t 0.4 mul sqrt  0.06}
	\parametricplot[plotstyle=dots,dotstyle=|,dotsize=0.2 0,plotpoints=28]%
		{1}{28}{t 2 mul sqrt  0.1}
	}
\psplot[plotstyle=curve,linecolor=blue,linewidth=2pt]{0.01}{14}{%
	x 8 div dup mul dup 0.01 exch div add %
	x 8 div dup mul 2.71828 exch exp 1 sub mul %
	0.5 exch div sqrt %
	}
\label{zones}
\refstepcounter{equation}
\rput[r](17,3){(\theequation)}
\end{pspicture}
$$

In the near zone $r\lesssim R_{\rm baryon}\ll (1/M_\Lambda)$,
the two baryons overlap and cannot be approximated as two separate instantons
of the $SU(2)$ gauge field;
instead, we need the ADHM solution of instanton number${}=2$ in all its complicated glory.
On the other hand,
{\it in the near zone, the nuclear force is five-dimensional:}
the curvature of the fifth dimension $z$ does not matter
at short distances, so we may treat the $U(2)$ gauge fields as living in
a flat 5D spacetime.
To leading order in $1/\lambda$, the $SU(2)$ fields are given by the ADHM solution,
while the abelian $\hat A_0(\vec x,z)$ is the 5D Coulomb field coupled
to the instanton density $(1/32\pi^2)\epsilon^{0KLMN}\tr(F_{KL}F_{MN})_{\rm ADHM}$.
Unfortunately, for two overlapping baryons this density has a rather complicated profile,
which makes calculating the near-zone nuclear force rather difficult.

The far zone $r\gtrsim(1/M_\Lambda)\gg R_{\rm baryon}$ poses the opposite problem:
The curvature of the 5D space and the $z$--dependence of the gauge coupling becomes
very important at large distances.
At the same time, the two baryons become well-separated instantons which may be treated
as point sources of the 5D abelian field $\hat A^0$.
In 4D terms, the baryons act as point sources for all the massive
vector mesons $A^\mu_n(x)$
comprising the massless 5D vector field $A^\mu(x,z)$, hence the nuclear force in the
far zone is the sum of 4D Yukawa forces,
\be
V(r)\ =\ {N_c^2\over4\kappa}\sum_n |\psi_n(z=0)|^2\times{e^{-m_n r}\over 4\pi r}
\label{FarYukawa}
\ee
where $m_n=O(M_\Lambda)$ are the vector meson's masses and $\psi_n(z)$
are their wave functions in the curved fifth dimension.
At the inner edge of the far zone, all the 4D vector mesons contribute to the
potential~(\ref{FarYukawa}) but for larger distances, the lightest vector meson
becomes dominant.

In the intermediate zone $R_{\rm baryon}\ll r\ll (1/M_\Lambda)$, we have the best of
both situations: The baryons do not overlap much {\sl and}
the fifth dimension is approximately flat.
At first blush, the nuclear force in this zone is simply the 5D Coulomb force between
two point sources,
\be
V(r)\ =\ {N_c^2\over 4\kappa}\times{1\over 4\pi^2 r^2}\
=\ {27\pi N_c\over 2\lambda M_\Lambda}\times{1\over r^2}\qquad
({\rm for}\ \zeta=1).
\label{FirstBlush}
\ee
This $1/r^2$ behavior of the repulsive potential suggests that the intermediate
zone of the holographic nuclear force corresponds to the repulsive hard core
of the real-life nucleons.

The real-life hard-core repulsion has both isoscalar and isovector components
of comparable strengths,
but the potential~(\ref{FirstBlush}) is purely isoscalar.
The reason for this discrepancy is that the point-source approximation
of holographic baryons is too crude for the intermediate zone where two
instantons of size $\rho=R_{\rm baryon}$ have $O((R_{\rm baryon}/r)^2)$
effects on each other.
Moreover, since the size of a stand-alone baryon is a compromise between two
sub-leading effects --- the $U(1)$ Coulomb repulsion and the $z$--dependence
of the gauge couplings $\kappa h(z)$ and $\kappa k(z)$ --- the baryons are linearly
sensitive to anything affecting  the leading $SU(2)$ gauge fields.
Hence, the overlap between two baryons gives rise to an
additional $1/r^2$ nuclear force of strength comparable to (\ref{FirstBlush}),
and since the overlap depends on the baryon's relative isospins,
this extra force has an isovector component.

To properly account for the baryon-baryon overlap,
Hashimoto, Sakai, and Sugimoto \cite{Hashimoto:2009ys}
(and also Kim and Zahed \cite {Kim:2008iy}) set the $SU(2)$ gauge fields to the
self-dual ADHM solution of instanton $\rm number=2$.
The instanton density $I(\vec x,z)=(1/32\pi^2)\epsilon^{0KLMN}\tr(F_{KL}F_{MN})_{\rm ADHM}$
of this solution deviates by $O((R_{\rm baryon}/r)^2)$ from the sum of
two separate instantons, and that has two effects:
(A) the $U(1)$ Coulomb energy is significantly different from~(\ref{FirstBlush}),
and (B) the width of the instanton density in $z$ direction is different,
which changes the $SU(2)$ field's energy since the gauge coupling depends on $z$.
Somehow, the two effects cancel out from the isoscalar components of the
hard-core potential, but they do give rise to isovector forces of comparable
magnitude.
Specifically, for baryons of $\rm spin=isospin={1\over2}$ ---
{\it i.e.,} for two nucleons --- Hashimoto {\it et~al} obtained
\be
V(r,{\bf n})\
=\ {27\pi N_c\over2\lambda M_\Lambda}\times{1\over r^2}\times\left(
	\hbox{\Large 1}_{\rm naive}\,
	+\,{64\over5}({\bf I}_1\cdot {\bf I}_2)({\bf n}\cdot{\bf J_1})({\bf n}\cdot{\bf J_2})
	\right)
\qquad\left({\bf n}={\vec r\over r}\,\right)
\label{HSS}
\ee
for the intermediate-zone distances $r$.
Note that the isoscalar component of this hard-core potential is precisely as
in the naive eq.~(\ref{FirstBlush}),
it's the isovector component that has really needed all the hard work.

\section{Attractive Forces in the Non-Antipodal Model}
\subsection{5D Scalars and their Interactions.}
In 4D, the attractive forces between two baryons emerge from exchanges of
virtual mesons with even spins and positive parity, especially the true scalars~$0^+$.
In the holographic theory, the baryons are instantons of the 5D non-abelian
gauge fields, while the 4D scalar mesons are modes of the 5D scalar field $\Phi(x)$,
so to get an attractive nuclear force we need a 5D scalar-vector coupling
of the form
\be
S_{\rm 5D}\ \supset\int\!\!d^5x\, \Phi\times\tr\bigl(F_{MN}^2\bigr).
\label{C5D}
\ee
In the Sakai--Sugimoto model, the 5D scalar $\Phi(x)$ describes deviations
of the D8 brane stack from its equilibrium position in the $(u,x^4)$ plane.\footnote{%
	In addition to the isosinglet scalar $\Phi(x)$ which describes the
	motion of the whole D-brane stack, there are also isotriplet scalar
	fields $\Phi^a(x)$ which describe the relative motion of the two D8 branes.
	For the moment, let us focus on the isosinglet $\Phi$, we shall
	return to the isotriplets later in this section.
	}
For the non-antipodal version of the model, such deviations have a $\delta u$
component, and since the local 5D gauge coupling depends on $u$,
we get $\delta u(\Phi)\times\tr(F_{MN}^2)$ interactions as in eq.~(\ref{C5D})
and hence the attractive force.
Unfortunately, for the antipodal model worked out by Sakai and Sugimoto themselves,
the equilibrium brane stack lies along the radius $x^4=\rm const$
(or rather two opposite radii), so the first-order deviations are in the $x^4$
direction only and have no $\delta u$ component.
Consequently, these is no linear $\Phi\times\tr(F^2_{MN})$ coupling in 5D
--- in fact, there is exact $Z_2$ symmetry $\Phi\to -\Phi$ which forbids it
--- and that's why there is no attractive nuclear force in the antipodal model.

In this section, we shall derive the effective 5D Lagrangian ---
including the crucial coupling (\ref{C5D}) --- for the non-antipodal model,
and then use it to derive the attractive nuclear force for the intermediate
distance.
Our first step is a precise definition of the scalar field $\Phi(x)$
in terms of the D8--brane stack deviation from its equilibrium position.
Using some kind of a non-singular coordinate $w$ along the brane stack,
we define:
\bea
{\rm In\ equilibrium:}\quad
u &=& \bar u(w),\qquad x^4\ =\ \bar x^4(w)
\label{equilibrium}\\
{\rm deviation:}\quad
u(w,x^\mu) &=&
\bar u(w)\,+\,\pi\alpha'\Phi(w,x^\mu),
\label{phidef}\\
x^4(w,x^\mu) &=&
\bar x^4(w)\,-\,\beta(w)\times\pi\alpha'\Phi(w,x^\mu),
\label{dev:x4}\\
{\rm where}\quad \beta(w) &=&
{g_{uu}(\bar u)\times(d\bar u/dw)\over g_{44}(\bar u)\times(d\bar x^4/dw)}\,.
\label{betadef}
\eea
The last formula here assures that to first order in $\Phi$,
the deviation of the stack is locally perpendicular to the stack itself.

In eqs.~(\ref{equilibrium}--\ref{betadef}) $w$ is a generic non-singular
coordinate along the un-perturbed brane stack, for example
$\bar z=[(\bar u^3-u_0^3)/u_0u_\Lambda^2]^{1/2}$ (but not the original $z$
which would be affected by the deviation field $\Phi$).
However, to simplify the 5D notations we would like to have the same metric
for all five dimension, $g_{ww}=g_{11}$, at least for $\Phi=0$.
This calls for
\be
dw^2\ =\ f(\bar u)(d\bar x^4(w))^2\
+\ {R_{D4}^3\over \bar u^3f(\bar u)}(d\bar u(w))^2\,,
\ee
which together with the brane equilibrium equation
\be
\left({d\bar x^4\over d\bar u}\right)^2\ =\ {R_{D4}^3\over \bar u^3f^2(\bar u)}\times
{u_0^8f(u_0)\over \bar u^8f(\bar u)-u_0^8f(u_0)}
\label{equeq}
\ee
gives us
\be
{d\bar u\over dw}\ =\ \sqrt{\bar u^8f(\bar u)-u_0^8f(u_0)\over \bar u^5R_{D4}^3}\,,\qquad
{d\bar x^4\over dw}\ =\ {u_0^4\sqrt{f(u_0)}\over \bar u^4f(\bar u)}\,,
\label{Weqs}
\ee
implicitly defining the $w$ coordinate.
We could not solve these equations analytically, but fortunately we would not need
the explicit formulae in this paper.
All we will need to know is that
\be
{\blue{\rm for\ small}\ w,\quad
\bar u(w)\ =\ \zeta u_\Lambda\left( 1\
	+\ {8\zeta^3-5\over 9\zeta^2}\times\bigl(M_\Lambda w\bigr)^2\
	+\ O\bigl((M_\Lambda w)^4\bigr)\right)}.
\label{Curvature}
\ee

For the above definitions, the 5D metric for $x^M=(x^\mu,w)$ becomes
\begin{align}
ds^2_{\rm 5d}\ =\ \left({\bar u+\pi\alpha'\Phi\over R}\right)^{3/2}\times
\biggl[ \eta_{MN}^{}\,dx^M dx^N\ &
+\ {\bar u^5 R_{D4}^3(\pi\alpha')^2\over u_0^8f(u_0)}\times(\partial_M\Phi\, dx^M)^2
\nonumber\\
&-\ 8{\pi\alpha'\Phi\over \bar u}\times dw^2\ +\ O(\Phi^2)\biggr],
\label{M5D}
\end{align}
while the $S^4$ radius and the dilaton depend on $\Phi$ according to
\be
{\rm radius}[S^4]\ =\ R_{D_4}^{3/4}(\bar u+\pi\alpha'\Phi)^{1/4},\qquad
e^\phi\ =\ g_s\left({\bar u+\pi\alpha'\Phi\over R_{D_4}}\right)^{3/4}.
\ee
Consequently, expanding the DBI action
\be
S_{\rm DBI}\
=\ T_8\!\int\!\!d^5x\,e^{-\phi}\mathop{\rm Vol}\nolimits(S^4)
\str\Bigl(\det\left(g_{MN}^{}+2\pi\alpha'\CF_{MN}\right)\Bigr)^{1/2}
\label{DBI}
\ee
to the second power in $\CF_{NM}$ and $\partial_M\Phi$ and to the first
power in $\Phi$ itself, we get
\bea
S_{\rm DBI} &=&
\!\!\int\!\!d^4x\,dw\Bigl( {\rm const}\,
	+\,{\cal L}_{\rm kin}\,+\,{\cal L}_{\rm int}\Bigr),
\label{SDBI}\\
{\cal L}_{\rm kin} &=&
{R_{D4}^3 \over 48\pi^4 g_s\ell_s^5}\left\{
	\bar u(w) \times\half\tr(\CF_{KL}^2)\,
	+\,{\bar u(w)^9\over u_0^8f(u_0)}\times\half(\partial_M\Phi)^2
	\right\}
\label{LK}\\
&=& {N_c\lambda M_{\Lambda}\zeta\over 216\pi^3}\left\{
	\bigl(\bar u(w)/u_0\bigr) \times\half\tr(\CF_{KL}^2)\,
	+\, {(\bar u(w)/u_0)^9\over 1-\zeta^{-3}}
		\times\half(\partial_M\Phi)^2\right\}
	\vrule width 0pt height 20pt depth 10pt\qquad\nonumber\\
&&\quad\langle\!\langle\, {\rm using\ the}\ \eta^{MN}\
	{\rm to\ contract\ the\ 5D\ indices,}\,
	\vrule width 0pt height 15 pt depth 10 pt\rangle\!\rangle\nonumber\\
{\cal L}_{\rm int} &=&
{N_c\over 48\pi^2}\Bigl(
	-3\Phi\times\half\tr(\CF_{\mu\nu}^2)\,
	+\,5\Phi\times\tr(\CF_{\mu w}^2)
	\Bigr)\
+\ O(\Phi^2).
\label{LI}
\eea

Thus far, we have focused only on the isosinglet scalar field $\Phi$
describing the common motion of the two flavor D8 branes
but ignored the isotriplet scalars $\Phi^a$ describing the relative
motion of the two D8 branes.
Fortunately, we may easily add the $\Phi^a$ to the 5D theory by
applying the $U(2)$ symmetry to the effective Lagrangian~(\ref{LK}--\ref{LI}).
Thus, without re-expanding the DBI action for two separated branes,
we  immediately obtain
\bea
{\cal L}_{\rm kin} &=&
{N_c\lambda M_{\Lambda}\zeta\over 216\pi^3}\biggl\{
	\bigl(\bar u(w)/u_0\bigr) \times
	\left(\tfrac14 (F_{MN}^a)^2\,+\,\tfrac14 \hat F_{MN}^2\right)
\label{Lkin}\\
&&\qquad\qquad\qquad+\,{(\bar u(w)/u_0)^9\over 1-\zeta^{-3}}\times
	\left(\tfrac12 (D_M\Phi^a)^2\,+\,\tfrac12(\partial_M\Phi)^2\right)
	\biggr\},
\nonumber\\
{\cal L}_{\rm int} &=&
{N_c\over 48\pi^2}\,\tr\Bigl( \bigl({\bf\Phi}=\Phi+\Phi^a\tau^a\bigr)\times
	\bigl(-\tfrac{3}{2}(\CF_{\mu\nu}^2)\,+\,5(\CF_{\mu w}^2)\bigr)\Bigr)\
	+\ O({\bf\Phi}^2)\nonumber\\
&=& {N_c\over 48\pi^2}\biggl\{
	\Phi\times\Bigl(-\tfrac{3}{4}(F_{\mu\nu}^a)^2\,
		+\,\tfrac{5}{2}(F_{\mu w}^a)^2\Bigr)\
	+\ \Phi\times\Bigl(-\tfrac{3}{4}(\hat F_{\mu\nu})^2\,
		+\,\tfrac{5}{2}(\hat F_{\mu w})^2\Bigr)
\vrule width 0pt height 20pt depth 10 pt \nonumber \\
&&\qquad\qquad\qquad+\ 2\Phi^a\times
	\left(-\tfrac{3}{4}\,F^a_{\mu\nu}\hat F_{\mu\nu}^{}\,
		+\,\tfrac{5}{2}\,F^a_{\mu w}\hat F^{}_{\mu w}\right)
	\biggr\}\ +\ O({\bf\Phi}^2).
\label{Lint}
\eea
However, for the two-baryon system we are interested in, the isotriplet
scalars fields $\Phi^a$ are much weaker than the isosinglet field $\Phi$
because they have much weaker sources.
Indeed, for two {\sl static} baryons, the $SU(2)$ gauge fields
are purely magnetic in 5D sense, {\it i.e.,} $F^a_{0i}=F^a_{0w}=0$,
while the $U(1)$ gauge fields are purely electric, $\hat F_{ij}=\hat F_{iw}=0$.
Consequently, on the last line of eq.~(\ref{Lint}), both
$F^a_{\mu\nu}\hat F^{}_{\mu\nu}=0$ and $F^a_{\mu w}\hat F^{}_{\mu w}=0$,
which leaves the isotriplet scalars $\Phi^a$ with no source at all.

For the baryons with non-zero spins, the $SU(2)$ electric fields do not
exactly vanish, but they are much weaker than their magnetic counterparts.
Specifically,
\be
{F_{\rm el}^{SU(2)}\over F_{\rm mag}^{SU(2)}}\
\sim\ \left({E_{\rm spin}\over E_{\rm static}}\right)^{1/2}\
\sim\ {J\over M_{\rm baryon}R_{\rm baryon}}\
\sim\ {1\over\sqrt{\lambda}}\times{J\over N_c}\
\ll\ 1.
\label{NAelectric}
\ee
At the same time, the abelian electric fields are also much weaker
then the non-abelian magnetic fields.
Indeed, were it not for the Chern--Simons interactions
\bea
{\cal L}_{\rm CS} &=&
{N_c\over 96\pi^2}\,\tr\Bigl(\epsilon^{JKLMN}{\cal A}_J\CF_{KL}\CF_{MN}+\cdots\Bigr)\nonumber\\
&=& {N_c\over 64\pi^2}\,\hat A_J^{}\times\epsilon^{JKLMN}
	\Bigl(\tr\bigl(F_{KL}F_{MN}\bigr)\,+\,\tfrac{1}{3}\hat F_{KL}\hat F_{MN}\Bigr)
\vrule width 0pt height 20pt depth 10pt\nonumber\\
&=& \tfrac12 N_c\hat A_J\times(\mbox{instanton current})^J\ +\ \cdots,
\label{CS}
\eea
the baryons would not generate any abelian fields at all.
As it is, for baryons of radius $\rho$, the non-abelian magnetic fields are
\be
F^{SU(2)}_{\rm mag}\ =
\begin{cases}
    O(1/\rho^2) & \text{in the near zone,}\\
	O(\rho/r^3) & \text{in the intermediate zone,}\\
\end{cases}
\ee
while the abelian electric fields are
\be
\hat F^{U(1)}_{\rm el}\ =
\begin{cases}
    O(N_c/\kappa \rho^3) & \text{in the near zone,}\\
	O(N_c/\kappa r^3) & \text{in the intermediate zone,}\\
\end{cases}
\ee
where $\kappa=O(N_c\lambda M_\Lambda)$ is the 5D kinetic-energy coefficient,
thus in both zones we have
\be
{\hat F^{U(1)}_{\rm el}\over F^{SU(2)}_{\rm mag}}\
\sim\ {N^c\over\kappa\rho}\ \sim\ {1\over \lambda M_\Lambda\rho}\
\sim\ {1\over\sqrt{\lambda}}\ \ll\ 1.
\ee
As to the abelian magnetic fields, they are generated by the Chern--Simons
terms involving $F^{SU(2)}_{\rm mag}\times F^{SU(2)}_{\rm el}$, so they
are even weaker than the electric abelian fields.
Altogether, we have a hierarchy of gauge fields
\be
F^{SU(2)}_{\rm mag}\ \ll\ F^{U(1)}_{\rm el}\
\ll\ F^{SU(2)}_{\rm el}\ \ll\ F^{U(1)}_{\rm mag}\,.
\label{Vhyerarchy}
\ee
Consequently,  the scalar-vector interaction Lagrangian~(\ref{Lint})
provides a much stronger source for the isosinglet scalar field $\Phi$
than to the isotriplet fields $\Phi^a$,
which leads to the scalar field hierarchy
\be
{\Phi^a\over\Phi}\ \sim\ {1\over\lambda}\times{J\over N_c}\ \ll\ 1.
\label{Shyerarchy}
\ee
Hence, the nuclear forces due to the triplet $\Phi^a$ are
much smaller then the forces due to the singlet $\Phi$,
so  we shall disregard the $\Phi^a$ through the rest of this article.

Focusing on the singlet scalar $\Phi$, we see that the dominant source
for it comes from the $SU(2)$ magnetic fields, so to the leading order in $1/\lambda$
we may approximate
\be
{\cal L}_{\rm int}\
\approx\ {N_c\over 48\pi^2}\,\Phi\times
\left(-\tfrac{3}{4}(F_{ij}^a)^2\,+\,\tfrac{5}{2}(F_{i w}^a)^2\right)\
+\ \text{unimportant.}
\ee
Moreover,  in the near and intermediate zones where the 5D gauge coupling
is approximately constant ---
$\kappa(w)=\kappa_0\times(\bar u(w)/u_0)\approx\kappa_0$ for $|w|\ll M_\Lambda^{-1}$,
{\it cf.}\ eq.~(\ref{Curvature}), ---
the $SU(2)$ fields are self-dual in the 4 space dimensions,
\be
F^a_{ij}(\vec x,w)\ =\ \epsilon_{ijk}F^a_{k w}(\vec x,w)
\label{SD}
\ee
Their specific form is given by the ADHM self-dual solution with
instanton $\rm number=2$, but fortunately we don't need the gory details
of this solution here.
All by itself, the self-duality assures us that
\be
\tfrac{1}{4}\bigl(F_{\mu\nu}^a\bigr)^2\
=\ \tfrac{1}{2}\bigl(F_{\mu w}^a\bigr)^2\\
=\ \tfrac{1}{16}\,\epsilon^{0KLMN}\bigl(F_{KL}^aF_{MN}^a\bigr)\
\equiv\ 4\pi^2\times{\rm instanton\ density}\ I(\vec x,w)
\ee
and hence, {\it\blue the scalar field $\Phi$ couples to the same instanton number
density as the abelian vector potential $\hat A_0$,}
\bea
{\cal L}_{\rm int} &=&
{N_c\over 48\pi^2}\,\Phi\times\bigl(-3\times 4\pi^2I\,+\,5\times 4\pi^2I\bigr)\
=\ {N_c\over 6}\,\Phi\times I,
\label{PhiI}\\
\text{cf.}\quad {\cal L}_{\rm SC} &=&
{N_c\over2}\,\hat A_0\times I\,.
\label{AI}
\eea
Therefore, in the near and intermediate zones where both the $\hat A^\mu$
and the $\Phi$ have approximately constant kinetic terms
\be
{\cal L}_{\rm kin}[\hat A_\mu,\Phi]\
\approx\ {N_c\lambda M_\Lambda\zeta\over 216\pi^3}\left\{
	\tfrac{1}{4}\hat F_{MN}^2\
	+\ {1\over 1-\zeta^{-3}}\times\tfrac{1}{2}(\partial_M\Phi)^2
	\right\},
\ee
both $\hat A^0$ and $\Phi$ are 5D Coulomb fields of the same charge density $I(\vec x,w)$,
\bea
\hat A_0(\vec x,w) &=&
{72\pi\over N_c\lambda M_\Lambda\zeta}\int\!\!d^3\vec x' dw'\,
{I(\vec x',w')\over (\vec x'-\vec x)^2+(w'-w)^2}\,,
\label{Coulomb5D}\\
\Phi(\vec x,w) &=& -{1-\zeta^{-3}\over 3}\times \hat A_0(\vec x,w).
\label{APhiRatio}
\eea
This fixed $\Phi/\hat A_0$ ratio has profound consequences for the attractive
nuclear force:
{\it\blue For any geometry of the instanton density $I(\vec x,w)$
in the near and intermediate zones,
there is a fixed ratio between the attractive force due to $\Phi$
and the repulsive force due to $\hat A_0$,} namely
\be
\red
{V^{\rm attractive}(r)\over V^{\rm repulsive}(r)}\
=\ -\rar\ =\ - {1-\zeta^{-3}\over 3^2}\ =\ \rm const.
\black
\label{MainResult}
\ee
Note that this ratio vanishes for $\zeta=1$ ---
there is no attractive scalar force in the antipodal model.
For the non-antipodal models, the attractive/repulsive force
ratio $\rar$ increases with $\zeta$, but it never gets larger than 1/9.
Thus, {\it in the near and intermediate zones of
the Sakai--Sugimoto model, the attractive nuclear force
is always weaker than the repulsive force.}

\subsection{Attractive Forces in the Intermediate Zone}
In light of eq.~(\ref{MainResult}), calculating the attractive
force between two nucleons at an intermediate distance
from each other
\be
\rho\,=\,R_{\rm baryon}\,\sim\,{\zeta^{-3/4}\lambda^{-1/2}\over M_\Lambda}\
\ll\ r\ \ll\ w_{\rm max}\,\sim\,{\zeta^{-1/2}\over M_\Lambda}
\ee
seems like a simple exercise.
Approximating both baryons as point sources of 5D Coulomb fields $\hat A_0$
and $\Phi$, we get
\bea
V^{\rm repulsive}(r) &=&
+{N_c^2\over 4\kappa(\zeta)}\times{1\over 4\pi^2 r^2}\
=\ +{27\pi N_c\over 2\zeta\lambda M_\Lambda}\times{1/r^2}\,,
\label{TrivRep}\\
V^{\rm attractive}(r) &=&
-{1-\zeta^{-3}\over 9}\times V^{\rm repulsive}(r)\
=\ -{3\pi N_c(1-\zeta^{-3})\over 2\zeta\lambda M_\Lambda}\times{1/r^2}\,,
\label{TrivAttr}
\eea
and the net potential is a repulsive $+1/r^2$ hard core.
Note that both the repulsive and the attractive potentials
(\ref{TrivRep}--\ref{TrivAttr}) are blind to spins and isospins
of the two baryons.
However, we saw in section~4 that such blindness is an artefact
of the treating the two baryons as point sources.
A better approximation makes the repulsive potential sensitive
to the baryon's spins and isospins, and we shall see momentarily
that the attractive potential has a similar sensitivity.

Indeed, let's follow  Hashimoto {\it et~al} \cite{Hashimoto:2009ys}
and take the $SU(2)$ gauge fields to be exactly self-dual, {\it i.e.,}
the two-instanton ADHM solution.
The instanton density of this solution
\be
I(\vec x,w)\ =\ {\epsilon^{0KLMN}\over 32\pi^2}\,\tr\Bigl(F_{LK}F_{MN}\Bigr)_{\rm ADHM}\
=\ I^{(1)}(\vec x,w)\ +\ I^{(2)}(\vec x,w)\ +\ \Delta I^{\rm overlap}(\vec x,w)\qquad
\label{Idensity}
\ee
differs from two separate instantons by $\Delta I^{\rm overlap}=O((\rho/r)^2)$,
which has two effects:
First, the repulsive Coulomb self-interaction of the instanton density~(\ref{Idensity})
\be
V[U(1)]\ =\ +{27\pi N_c\over 2\zeta\lambda M_\Lambda}\times\left\{
	{1\over r^2}\
	+\int\!\!d^3\vec x\,dw\,{\Delta I^{\rm overlap}(\vec x,w)\over (\vec x-\vec X_1)^2+w^2}\
	+\ \left(\vec X_1\to\vec X_2\right)\
	+\ O\left((\Delta I)^2\right)
	\right\}
\label{NonTrivRep}
\ee
is more complicated then~(\ref{TrivRep}), and second,
for a $w$--dependent gauge coupling $\kappa(w)=\kappa(0)\times(\bar u(w)/u_0)$
--- {\it cf.}\ eqs.~(\ref{Lkin}) and (\ref{Curvature}) ---
re-distribution of the $SU(2)$ gauge fields in the $w$ direction changes
their energy by
\bea
\Delta E[SU(2)] &=&
{N_c\lambda M_\Lambda\zeta\over 216\pi^3}\left\{
	\int\!\!d^3\vec x\,dw\bigl(\bar u(w)/u_0\bigr)\times\tfrac12\tr\bigl(F^2_{MN}\bigr)\
	-\ 2\times 8\pi^2\right\}
\nonumber\\
&=&\ {N_c\lambda M_\Lambda\over 81\pi}\times{8\zeta^3-5\over 3\zeta}\times
\int\!\!d^3\vec x\,dw\,w^2\times\Delta I^{\rm overlap}(\vec x,w).
\label{DEsu2}
\eea
Both effects were evaluated in careful detail by
Hashimoto {\it et~al} \cite{Hashimoto:2009ys}, so let us simply adapt their
results to the present situation.
For two {\sl nucleons --- \it i.e.,} baryons of $\rm spin=isospin={1\over2}$ ---
and $N_c\gg1$ (which suppresses the quantum fluctuations of the baryons' sizes
or locations), they found
\bea
\Delta E[SU(2)] &=&
+{A\over r^2}\times\left( 1\ +\
	{16\over9}\bigl({\bf I}_1\cdot {\bf I}_2\bigr)\times
	\left[ 2\bigl({\bf n}\cdot{\bf J}_1\bigr)\bigl({\bf n}\cdot{\bf J}_2\bigr)\,
		-\,\bigl({\bf J}_1\cdot{\bf J}_2\bigr)\right]
	\right)
\label{Esu2}\\
{\rm where}\ A &=&
{\lambda N_c M^3_\Lambda\over 162\pi}\times{8\zeta^3-5\over 3\zeta}\times\rho^4,
\label{Aeq}\\
V[U(1)] &=&
+{B\over r^2}\times\left(
	\hbox{\Large 1}_{\rm naive}\ -\ {2\over 5}\
	+\ {32\over 45}\bigl({\bf I}_1\cdot {\bf I}_2\bigr)\times
		\bigl({\bf J}_1\cdot{\bf J}_2\bigr)
	\right)
\label{Vu1}\\
{\rm where}\ B &=&
{27\pi N_c\over 2\zeta\lambda M_\Lambda}\,.
\label{Beq}
\eea
Or rather, they obtained these formulae for the antipodal modal, but the extra
$\zeta$--dependent factors for the non-antipodal models are obvious from
eqs.~(\ref{DEsu2}) and (\ref{NonTrivRep}).

Now consider the attractive scalar force between two baryons.
Since the scalar field $\Phi$ couples to the same non-trivial instanton density
(\ref{Idensity}) as the abelian gauge field $\hat A_0$, the attractive force
has the same complicated spin and isospin dependence as the $U(1)$ force~(\ref{Vu1}):
Instead of the naive eq.~(\ref{TrivAttr}), we get
\be
V[\Phi]\ =\ -\rar\times V[U(1)]\
=\ -{\rar B\over r^2}\times\left(
	\hbox{\Large 1}_{\rm naive}\ -\ {2\over 5}\
	+\ {32\over 45}\bigl({\bf I}_1\cdot {\bf I}_2\bigr)\times
		\bigl({\bf J}_1\cdot{\bf J}_2\bigr)
	\right)
\label{Vphi}
\ee
where $\rar=\frac19 (1-\zeta^{-3})$ as in eq.~(\ref{MainResult}).

But besides the direct contribution~(\ref{Vphi}) of the scalar force
to the two-baryon potential, it also affects the baryon radius $\rho$
--- which in turn affects the $SU(2)$
force according to eqs.~(\ref{Esu2}--\ref{Aeq}).
To see how this works, consider a stand-alone $SU(2)$ instanton with
\be
I(\vec x,w)\ =\ {6\rho^4\over\pi(\vec x^2+w^2+\rho^2)^4}\,.
\ee
The classical energy of this instanton is
\be
E(\rho)\ =\ E[SU(2)]\ +\ \Delta E[U(1)]\ +\ \Delta E[\Phi]
\ee
where
\be
E[SU(2)]\ =\ {N_c\lambda M_\Lambda\zeta\over 27\pi}\int\!\!d^3\vec x\,dw\,
I(\vec x,w)\times{\bar u(w)\over u_0}\
=\ {N_c\lambda M_\Lambda\zeta\over 27\pi}\
+\ {\lambda N_c M^3_\Lambda\over 162\pi}\times{8\zeta^3-5\over 3\zeta}\times\rho^2,
\ee
the $\hat A_0$--mediated Coulomb self-interaction is
\be
\Delta E[U(1)]\ =\ +B\times{2\over 5\rho^2}
\ee
where $B$ is as in eq.~(\ref{Beq}),
and the $\Phi$--mediated Coulomb self-interaction is
\be
\Delta E[\Phi]\ =\ -\rar\times\Delta E[U(1)]\
=\ -\rar B\times{2\over 5\rho^2}\,.
\ee
Note that the scalar / abelian vector force ratio~(\ref{MainResult}) works at short
distances as well as intermediate.

Assembling all the contributions to the classical instanton energy, we get
\be
E(\rho)\ =\ {N_c\lambda M_\Lambda\zeta\over 27\pi}\
+\ {\lambda N_c M^3_\Lambda\over 162\pi}\times{8\zeta^3-5\over 3\zeta}\times\rho^2\
+\ {27\pi N_c\over 5\lambda M_\Lambda}\times{1-\rar\over\zeta}\times{1\over\rho^2}\,,
\ee
and minimizing this formula with respect to $\rho$ gives us the classical mass and
radius of the baryon:
\bea
R_{\rm baryon}\ =\ \rho_{\rm @min} &=&
\left({81\pi\sqrt{2/15}\over\lambda M_\Lambda^2}\right)^{1/2}
	\times\left({3\over 8\zeta^3-5}\right)^{1/4}
	{\red\times\left(1-\rar\right)^{1/4}},
\label{Rbaryon}\\
M_{\rm baryon} &=&
{N_c\lambda M_\Lambda\zeta\over 27\pi}\
	+\ {M_\Lambda\over \sqrt{30}}\left({8\zeta^3-5\over 3\zeta^2}\right)^{1/2}
	{\red\times\left(1-\rar\right)^{1/2}}\,,
\vrule width 0pt height 20pt
\label{Mbaryon}
\eea
Note that these formulae differ from eqs.~(\ref{BMR}) by factors
$(1-\rar)^{\rm some\,power}$: These factors are due to $\Phi$--mediated
attractive force which we didn't take into account back in section~3.
In particular, the scalar force reduces the baryon radius by a factor
$(1-\rar)^{1/4}$.
Substituting this radius into eq.~(\ref{Aeq}) gives us
\be
A\ =\ {2B\over5}{\red \times\left(1-\rar\right)},
\ee
which means that  the indirect effect of the scalar force
reduces the $SU(2)$-mediated nuclear force by the factor $(1-\rar)$.
At the same time, the direct effect~(\ref{Vphi}) of the scalar force
reduces the net Coulomb force by exactly the same factor,
\be
V[U(1)]\ +\ V[\Phi]\ =\ V[U(1)]{\red \times\left(1-\rar\right)}.
\ee
Altogether, we end up with this $(1-\rar)$ factor multiplying the whole
nuclear force in all its complicated glory,
\bea
V_{\rm net}(r) &=&
\Delta E[SU(2)]\ +\ V[U(1)]\ +\ V[\Phi]
\nonumber\\
&=& {2B\over5r^2}{\red \times\left(1-\rar\right)}\times\left( 1\ +\
	{16\over9}\bigl({\bf I}_1\cdot {\bf I}_2\bigr)\times
	\Bigl[ 2\bigl({\bf n}\cdot{\bf J}_1\bigr)\bigl({\bf n}\cdot{\bf J}_2\bigr)\,
		-\,\bigl({\bf J}_1\cdot{\bf J}_2\bigr)\Bigr]
	\right)
\nonumber\\
&&\qquad+\ {B\over r^2}{\red \times\left(1-\rar\right)}\times\left(
	\hbox{\Large 1}_{\rm naive}\ -\ {2\over 5}\
	+\ {32\over 45}\bigl({\bf I}_1\cdot {\bf I}_2\bigr)\times
		\bigl({\bf J}_1\cdot{\bf J}_2\bigr)
	\right)
\nonumber\\
&=& {B\over r^2}{\red \times\left(1-\rar\right)}\times\left(
	\hbox{\Large 1}_{\rm naive}\
	+\ {64\over 45}\bigl({\bf I}_1\cdot {\bf I}_2\bigr)\times
		\bigl({\bf n}\cdot{\bf J}_1\bigr)\bigl({\bf n}\cdot{\bf J}_2\bigr)
	\right).
\label{NetForce}
\eea

To summarize, {\blue the net holographic nuclear force in the intermediate zone
is precisely as in Hashimoto, Sakai, and Sugimoto \cite{Hashimoto:2009ys}.
The only effect of the attractive force due to the scalar field $\Phi$
is to reduce the whole force by a constant overall factor $(1-\rar)$.
Depending on the $\zeta$ parameter of the model, this factor varies between
${8\over9}$ (reduction by 11\%) and 1 (no reduction at all), but it is always positive
so the net force in the intermediate zone is always repulsive.}

\subsection{Attractive Forces in the Far Zone}
In the far zone, the $w$--dependence of the gauge coupling
\be
\kappa(w)\ =\ {N_c \lambda M_\Lambda\zeta\over 216\pi^3}\times{\bar u(w)\over u_0}
\ee
becomes important and the lowest-energy two-instanton solution of the
$SU(2)$ equation $D_M(\kappa F^{MN})=0$ is no longer self-dual.
Consequently, working out the overlap between two distant instantons
becomes rather difficult and we are reduced to a cruder point-source
approximation.
Earlier, we saw that in the intermediate zone, this approximation has yielded
a correct isoscalar central force but missed
the isovector spin-spin and tensor forces.
By analogy, we expect that in the far zone, the point-source approximation
would give us a correct isoscalar force, but the isovector forces could be wrong.

Let's focus on the isoscalar forces.
In the point-source approximation, they follow the Green's functions
of the $\hat A_0(\vec x,w)$ and $\Phi(\vec x,w)$ fields in the far zone,
\bea
V^{\rm repulsive}(r) &=&
+{54\pi^3 N_c\over\zeta\lambda M_\Lambda}\times
\label{VRFZ}\\
&&\qquad\times\bra{\vec x,0\strut}\left[
	\left({\bar u(w)\over u_0}\right)\vec\nabla^2\,
	+\,{\partial\over\partial w}\left({\bar u(w)\over u_0}\right){\partial\over\partial w}
	\right]^{-1}\ket{\vec 0,0},\nonumber\\
V^{\rm attractive}(r) &=&
-\rar\times{54\pi^3 N_c\vrule width 0pt height 20pt\over\zeta\lambda M_\Lambda}\times
\label{VAFZ}\\
&&\qquad\times\bra{\vec x,0\strut}\left[
	\left({\bar u(w)\over u_0}\right)^9\vec\nabla^2\,
	+\,{\partial\over\partial w}\left({\bar u(w)\over u_0}\right)^9{\partial\over\partial w}\,
	+\,\mu^2(w)
	\right]^{-1}\ket{\vec 0,0},\nonumber
\eea
where
\be
\mu^2(w)\ =\ -{2\zeta M^2_\Lambda\over9}\times\left[
	35\left({\bar u(w)\over u_0}\right)^{10}\,
	-\,32\zeta^{-3}\left({\bar u(w)\over u_0}\right)^7\,
	+\,13(1-\zeta^{-3})\left({\bar u(w)\over u_0}\right)^2
	\right]
\label{MU2}
\ee
is the effective mass term for the $\Phi$ in the expansion of the DBI Lagrangian
in terms of 5D coordinate $(x^\mu,w)$.\footnote{%
	The negative sign of $\mu^2(w)$ is an artefact of the $w$--dependent
	normalization of the scalar field;
	a canonically normalized $\Phi$ has positive $\rm mass^2$
	\be
	\mu_{\rm can}^2(w)\ =\ +{\zeta M_\Lambda^2\over 9}\times\left[
		20\left({\bar u(w)\over u_0}\right)\,
		+\,\zeta^{-3}\left({\bar u(w)\over u_0}\right)^{-2}\,
		-\,44(1-\zeta^{-3})\left({\bar u(w)\over u_0}\right)^{-7}
	\right].
	\ee
	}
Expanding the 5D fields $\hat A_0(\vec x,w)$ and $\Phi(\vec x,w)$
in terms of the 4D vector and scalar mesons, we may express their Green's functions
--- and hence the potentials~(\ref{VRFZ}--\ref{VAFZ}) ---
as sums of Yukawa potentials
\bea
V^{\rm repulsive}(r) &=&
+{27\pi^2 N_c\over2\zeta\lambda M_\Lambda}\times
\sum_{n=1}^\infty\left|\Psi_n^V(w=0)\right|^2\times{\exp(-m_n^V r)\over r}\,,
\label{VRY}\\
V^{\rm attractive}(r) &=&
-\rar\times{27\pi^2 N_c\over2\zeta\lambda M_\Lambda}\times
\sum_{n=1}^\infty\left|\Psi_n^S(w=0)\right|^2\times{\exp(-m_n^S r)\over r}\,.
\label{VAY}
\eea
Here $\Psi_n^V(w)$ and $\Psi_n^S(w)$ are the wave functions in the fifth
dimension of the respective vector or scalar mesons.
Note that only the odd--$n$ modes --- which give rise to the
true vector $1^-$ and true scalar $0^+$ mesons ---
contribute to the sums~(\ref{VRY}--\ref{VAY}).
The even--$n$ modes --- responsible for the
axial-vector $1^+$ and pseudo-scalar $0^-$ mesons ---
don't contribute because their wave functions vanish at $w=0$.

We do not have analytical formulae for the meson's masses or wave functions,
but numerical calculations \cite{Sakai:2004cn,Mintakevich:2008mm} show that
for Sakai--Sugimoto models with any $\zeta\ge1$, the vector mesons are always
lighter than the scalar mesons,
\be
\forall n,\quad m_n^V\ <\ m_n^S.
\ee
Numerical calculation of the mesons' wave functions at $w=0$ is still in progress,
but there does not seem to be much difference between vector and scalar mesons,
so it is reasonable to {\it assume} that
\be
{\rm for\ odd}\ n,\quad{\left|\Psi_n^V(0)\right|^2\over\left|\Psi_n^S(0)\right|^2}\
>\ \rar\ =\ {1-\zeta^{-3}\over 9}\,.
\label{Assumption}
\ee
If this assumption is correct, then every term in the repulsive potential~(\ref{VRY})
is stronger than the corresponding term in the attractive potential~(\ref{VAY}), and
{\it the net isoscalar force is repulsive throughout the far zone.}

At the outer end of the far zone, we don't need the assumption~(\ref{Assumption})
to show that the repulsion is stronger than attraction, all we need to know
is that the lightest vector meson is lighter than the lightest scalar meson,
$m_1^V< m_1^S$.
Indeed, for $r\gg (1/m_{\rm meson})\sim(1/M_\Lambda\sqrt\zeta)$,
each sum  (\ref{VRY}--\ref{VAY}) is dominated by the slowest-decaying Yukawa term
belonging to the lightest vector or scalar meson,
\be
V^{\rm repulsive}(r)\ \sim\
+O(N_c/\lambda)\times{\exp(-m_1^V r)\over r}\,,\qquad
V^{\rm attractive}(r)\ \sim\
-O(N_c/\lambda)\times{\exp(-m_1^S r)\over r}\,.
\ee
Regardless of the pre-exponential factors here, the longer-ranged force
always wins over the shorter-ranged force at long distances,
and since $m_1^V<m_1^S$ in all versions of the Sakai--Sugimoto model
--- antipodal and non-antipodal with any $\zeta>1$ ---
{\blue the net {\it isoscalar} force is repulsive at long distances
$r\gg 1/(M_\Lambda\sqrt{\zeta})$.}

Note that this behavior of the Sakai--Sugimoto model is very different from
the real-life nuclear physics.
Indeed, in reality
the lightest scalar meson $\sigma$ is lighter then the lightest vector meson,
$M_\sigma\sim 600$~MeV while $M_\omega\approx 787$~MeV, and consequently
{\it the real-life isoscalar nuclear force is attractive rather than
repulsive at short distances.}
We do not know the reason for this discrepancy;
perhaps it's a peculiar bad feature of the Sakai--Sugimoto scheme and could
be fixed by an alternative holographic model.
But perhaps it's an inherent problem of the large--$N_c$ nuclear physics.
As explained in section~2, the QCD origin of the $\sigma(600)$ resonance
is rather controversial --- maybe it's a true meson originating in the
$\sigma$ field of the linear-sigma-model-like chiral symmetry breaking, or maybe it's
just a two-pion resonance which would not exist without strong $\pi\pi$ interactions.
In the first scenario, the large $N_c$ limit would make
$\sigma(600)$ into a narrow resonance, but it would still be there and lighter than
the lightest vector meson $\omega$, so the attractive nuclear force would have a longer range
than the repulsive force.
But in the second scenario, there would be no $\sigma(600)$ resonance in the
large $N_c$ limit, the lightest remaining scalar meson\footnote{%
	Probably the  $f_0(980)$, or maybe even the
	$f_0(1450)$ --- which is the best fit for the lightest scalar
	in the Sakai--Sugimoto model --- in case the $f_0(980)$ is also a two-pion
	resonance that would go away when $N_c\to\infty$.
	Either way, the lightest surviving scalar would be heavier than
	$\omega(787)$.
	}
would be heavier than $\omega$, and the dominant
isoscalar nuclear force at long distances would be repulsive rather than
attractive.

The best way to settle this issue would be to find the $\sigma$ resonance
and its mass in a lattice QCD calculation for several values of $N_c$, which
would hopefully allow us to extrapolate to $N_c\to\infty$.
Alternatively, once we have several different holographic models, we can compare
their predictions for the meson spectra in general and for the lightest scalar
to lightest vector mass ratio in particular.
Either way, this issue will have to wait for future research.

Another issue that would have to wait for future research involves
nuclear forces arising from nucleons exchanging pairs of un-bound mesons
--- especially pairs of pions --- rather than single mesons or resonances.
In the real-life nuclear physics, the double pion exchange generates
the longest-ranging attractive isoscalar force,
which makes a significant contribution to the bulk binding energy of the nuclei.
In the large $N_c$ limit, the double-pion exchange decreases as $1/N_c$
relatively to the single-meson exchanges, but it would remain
significant at longer distances where the $\exp(-mr)$ factors for single mesons
like $\sigma$ or $\omega$ are even smaller than $1/N_c$.
Unfortunately, in holographic duals of QCD,  the double-meson exchanges happen
at the one-string-loop level, so they cannot be
be calculated in terms of an effective semi-classical 5D gravity
--- or even 10D gravity plus other local fields --- but require a fully quantum
string theory.
Since string perturbation theory in curved backgrounds is rather hard,
we leave the double-meson exchanges for future research.

Instead, lets us now address the dominant nuclear force at the longest distances,
namely the isovector force due to single pion exchange between two nucleons,
\bea
V^\pi(r) &=&
-{g_A^2\over\pi f_\pi^2}\,\bigl({\bf I_1}{\bf I}_2\bigr)\left[
	\bigl({\bf J}_1{\bf J}_2\bigr)\times{m_\pi^2\over 3r}\,
	+\,T_{12}({\bf n})\times
	\left({m_\pi^2\over 3r}+{m_\pi\over r^2}+{1\over r^3}\right)
	\right]e^{-m_\pi r}\nonumber\\
& \setbox0=\hbox{$\scriptstyle{\,m_\pi\to0\,}$}
	\mathop{\hbox to \wd0 {\rightarrowfill}}\limits_{m_\pi\to0} &
-{g_A^2\over\pi f_\pi^2}\,\bigl({\bf I_1}{\bf I}_2\bigr)T_{12}({\bf n})
\times{1\over r^3}
\eea
where $T_{12}({\bf n})=3({\bf n} {\bf J_1})({\bf n} {\bf J_2})
-({\bf J}_1{\bf J}_2)$ is the direction dependence of the tensor force.
The overall coefficient of this force was calculated by Hashimoto {\it et~al}
\cite{Hashimoto:2008zw,Hashimoto:2009ys} for the antipodal model.
Adapting their method to non-antipodal models gives us
\be
{g_A^2\over \pi f_\pi^2}\
=\ {8N_c\lambda\zeta^{3/2}M_\Lambda^2\over 3^6\pi\,F(\zeta)}\times\rho^4
\label{PIforce}
\ee
where $\rho$ is the classical radius of the baryon and $F^{-1/2}$
is the normalization factor of the pion's wave function
\bea
\Psi_\pi(w) &=&
{1\over\sqrt{\kappa_0(\zeta) F(\zeta)}}\times{u_0\over\bar u(w)}\,,\\
F(\zeta) &=&
{M\sqrt\zeta\over3}\int\limits_{-w_{\rm max}}^{+w_{\rm max}}\!\!dw\,{u_0\over\bar u(w)}\
=\int\limits_1^\infty\!\!dy\left({y^3\over y^8-\zeta^{-3}y^5-(1-\zeta^{-3})}\right)^{1/2},\\
&&{\pi\over 3}\,<\,F(\zeta)\,<\,{\sqrt\pi\Gamma(3/16)\over 4\Gamma(11/16)}\,\approx\,1.65.
\eea
The pions themselves are modes of the $SU(2)$ gauge fields $A^a_M(x,w)$
and don't know the 5D scalar field $\Phi(x,w)$ from the hole in the ground.
However, their interactions with baryons depend on the classical baryon radius~$\rho$,
and because of the $\Phi$--mediated attractive forces {\sl in the near zone,}
this radius is smaller than it would have been otherwise,
$$
\rho^4\
=\ \left({81\pi\sqrt{2/15}\over\lambda M_\Lambda^2}\right)^2
	\times\left({3\over 8\zeta^3-5}\right)
	{\red\times\left(1-\rar\right)}.
\eqno(\ref{Rbaryon})
$$
Consequently, {\blue the pion-mediated nuclear force in the far zone
becomes sensitive to the $\Phi$-mediated attractive force in the near zone,}
\be
V^\pi(r)\
=\ -{({\bf I_1}{\bf I}_2)T_{12}({\bf n})\over r^3}\times
{48 N_c\over 5\lambda M_\Lambda^2}\times
{\pi\zeta^{3/2}\over(8\zeta^3-5)F(\zeta)}
{\red\times\left(1-\rar\right)}.
\ee

What about the other isovector mesons' contributions to the nuclear force?
In the Sakai--Sugimoto model we have (pseudo) scalar isovector mesons
coming from the modes of the 5D scalar fields $\Phi^a(x,w)$, but their
couplings to baryons are too small to create an appreciable force.
In addition, we have vector and axial vector isovector mesons $\rho$ (770~MeV),
$a_1$ (1260~MeV), {\it etc.,} coming from the non-zero modes of the $SU(2)$
gauge fields $A_M^a(x,w)$.
Their contributions to the isovector spin-spin and tensor forces were
calculated by Hashimoto~{\it et~al}~\cite{Hashimoto:2009ys} as
\be
V^{\rm other}_{\rm isovector}(r)\
=\ -{128\pi^3\over27}\,\kappa_0^2\,{\red\rho^4}\,({\bf I}_1{\bf I_2})
\sum_{n=1}^\infty Q_n e^{-m_n r}\left[
	({\bf J_1}{\bf J_2})\,{m_n^2\over 3r}\,
	+\,T_{12}({\bf n})\left({1\over r^3}+{m_n\over r^2}+{m_n^2\over 3r}\right)
	\right]
\ee
where
\be
Q_n\ =\begin{cases}
	+\left|\Psi_n(w=0)\right|^2 & {\rm for\ odd}\ n,\\
	-{1\over m_n^2}\left|\Psi_n'(w=0)\right|^2 & {\rm for\ even}\ n.\\
	\end{cases}
\ee
Note the overall factor $\red\rho^4$ for all the isovector forces.
Since the baryon radius is affected by the attractive forces in the near zone,
they have indirect effect on all the isovector forces in the far zone,
\be
V^{\rm isovector}(r)\ \propto {\red\left(1-\rar\right)}.
\ee
This rule applies even to the effects of the baryon-baryon overlap
that Hashimoto {\it et~al} could not calculate in the far zone.
We cannot calculate their $r$ dependence either, but their $\rho$
dependence should be the same as in the intermediate zone,
\be
V^{\rm overlap}(r)\ \propto\ \rho^4\ \propto {\red\left(1-\rar\right)}.
\ee

%
\section{Full DBI Action in the Near Zone}
In the previous section we studied nuclear forces in the
intermediate and far zones, but since those forces depend on the baryon
radius~$\rho$, we had to stick our noses into the near zone to see how
$\rho$ is affected by the scalar-mediated forces.
In the intermediate and far zones, all the gauge fields are weak enough
so we could expand the DBI Lagrangian in power of $\CF_{MN}$ and truncate
the expansion after the leading Yang--Mills term as we did in
eqs.~(\ref{SDBI}--\ref{LI}),
\bea
{\rm for}\ r\,\gg\,\rho\,\sim\,{1\over\sqrt\lambda M_\Lambda}\,,\quad
2\pi\alpha'\CF_{MN}&\ll&G_{MN}\nonumber\\
&\hbox{\hskip 0.25em\rput{270}(0,0){$\Longrightarrow$}\hskip 0.25em}&
\label{YMapprox}\\
{\cal L}_{\rm DBI}\ \propto\ \str\sqrt{\det(g+2\pi\alpha'\CF)}
&\approx& \sqrt{-\det(g)}\times\left(
	2\,+\,{(2\pi\alpha')^2\over4}\tr\left(\CF_{MN}\CF^{MN}\right)\right).\qquad\nonumber
\eea
However, in the near zone $r\sim\rho$ the $SU(2)$ gauge fields become too strong
for the YM approximation,
\be
{\rm for}\ r\sim\rho\quad2\pi\alpha'\CF_{MN}\,\sim\, g_{MN}\quad\Longrightarrow\quad
{\cal L}_{\rm DBI}\ \not\approx\
\sqrt{-\det(g)}\left(2\,+\,{(2\pi\alpha')^2\over4}\tr\left(\CF_{MN}\CF^{MN}\right)\right),
\label{BigFields}
\ee
which casts doubt on accuracy of eqs.~(\ref{Rbaryon}--\ref{Mbaryon})
for the baryon's mass and radius and hence of all the $\rho$--dependent
formulae for isovector nuclear forces.

In this section we shall see that despite the higher-order DBI interactions beyond the
Yang--Mills approximation, to leading order in $1/\lambda$ the baryon radius
remains exactly as in eq.~(\ref{Rbaryon}) and we do not need to make
any leading-order corrections to the nuclear forces we have computed
in the previous section~4.
And at this happy note, the readers impatient with technical details may skip the
rest of this section and go straight to the summary section~7.

Our main point is that while the Yang-Mills approximation~(\ref{YMapprox})
to the full DBI action\footnote{%
	Actually, the $\rm DBI+CS$ action is also incomplete --- for multiple D-branes,
	there are additional terms involving covariant derivatives ${\cal D}_L\CF_{MN}$
	of the non-abelian gauge fields.
	Fortunately, for instanton-like gauge fields of a baryon, such covariant
	derivatives are relatively small --- even when the gauge fields themselves
	are large as in eq.~(\ref{BigFields}) ---
	so following Tseytlin \cite{Tseytlin:1997csa} we shall limit our analysis
	to the non-abelian DBI action.
	}
does not work for {\it generic} strong gauge fields,
it may work `by accident' for some special $\CF_{MN}(x)$ configurations.
In particular, for self-dual gauge fields living on a flat D4 brane stack,
the YM approximation not only
works but happens to be exact, regardless of the number of instantons or their sizes
\cite{Brecher:1998su},
\be
{\rm for}\ \CF_{\mu\nu}\,
=\,\tfrac12\epsilon_{\mu\nu\alpha\beta}\CF^{\alpha\beta},\quad
\str\sqrt{\det({\bf1}_{4\times4}+2\pi\alpha'\CF)}\
=\ 2\ +\ {(2\pi\alpha')^2\over4}\tr\left(\CF_{MN}\CF^{MN}\right)
\label{DBIYM}
\ee
In our case, the 5D metric is not flat, and besides the (approximately)
self-dual $SU(2)$ magnetic fields we also have the abelian electric field.
Nevertheless, eq.~(\ref{DBIYM}) continues to hold exactly for a slightly
modified self-duality condition for the $SU(2)$ magnetic fields.

Indeed, consider a more general case of some non-flat but static metric
\be
ds^2\ =\ -|g_{00}(x)|dt^2\ +\ g_{mn}(x)dx^m dx^n\qquad (m,n=1,2,3,4),
\ee
arbitrary but purely-electric abelian fields $\hat F_{m0}(x)$ and purely-magnetic
$SU(2)$ fields $F^a_{mn}(x)$.
In general, Tseytlin's non-abelian version \cite{Tseytlin:1997csa}
of the DBI Lagrangian works like this:
First, one calculates the determinant of the
 $g_{MN}+2\pi\alpha'\CF_{MN}$ matrix (in space-time indices) while completely ignoring
their gauge indices or the fact that they don't commute with each other.
Second, one expands the square root of this determinant into a formal power series
in the gauge fields.
Finally, for each term in this determinant one takes a {\it symmetrized trace}
over the gauge indices, and then tries to re-sum the series.
For the static case at hand we have
\be
g_{MN}\,+\,2\pi\alpha'\CF_{MN}\ =\,\begin{pmatrix}
	-|g_{00}| &\vrule& -\pi\alpha'\hat F_{m0}\\
	\noalign{\hrule}
	+\pi\alpha'\hat F_{0m} &\vrule& g_{mn}+2\pi\alpha'F_{mn}\\
	\end{pmatrix}
\ee
and hence
\be
-\det\limits_{\rm 5D}\bigl(g_{MN}+2\pi\alpha'\CF_{MN}\bigr)\
=\ |g_{00}|\times\det\limits_{\rm 4D}\bigl(\hat g_{mn}+2\pi\alpha'F_{mn}\bigr)
\label{determinants}
\ee
where
\be
\hat g_{mn}(x)\ =\ g_{mn}(x)\ -\ (\pi\alpha')^2|g^{00}|\,\hat F_{m0}\hat F_{n0}\,.
\label{hatmet}
\ee
Note that while the determinants on both sides of eq.~(\ref{determinants})
are formal --- they ignore the non-commutativity of the magnetic fields
$F_{mn}=\tfrac12\tau^a F^a_{mn}$ --- but the modified metric~(\ref{hatmet})
does not have any non-commutativity problems because the electric fields
$\hat F_{m0}$ are purely abelian.
When those electric fields are too strong, we can get a different problem of
$\hat g_{mn}$ matrix loosing positive-definiteness, but fortunately
the $U(1)$ fields of a baryon never get that strong: Even in the near zone,
\be
(\pi\alpha')^2|g^{00}|g^{mn}\hat F_{m0}\hat F_{n0}\
\lesssim\ O(1/\lambda)\ \ll\ 1,
\ee
so the modified metric $\hat g_{mn}$ remains safely positive-definite.

On the right hand side of eq.~(\ref{determinants}), the 4D determinant evaluates to
\be
\det\bigl(\hat g_{mn}+2\pi\alpha'F_{mn}\bigr)\
=\ \det(\hat g_{mn})\times\left[ 1\,
	+\,2(\pi\alpha')^2\hat g^{mp}\hat g^{nq}F_{mn}F_{pq}\,
	+\,(\pi\alpha')^4\left(\hat g^{mp}\hat g^{nq}F_{mn}\tilde F_{pq}\right)^2
	\right]
\label{det4D}
\ee
where
\be
\tilde F_{pq}\ =\ \tfrac12\sqrt{\det(\hat g)}\epsilon_{pqrs}
\hat g^{rm}\hat g^{sn} F_{mn}
\label{MHD}
\ee
 is the Hodge dual of the $F_{mn}$ {\sl with respect
to the modified metric}~$\hat g_{mn}$.
When the $SU(2)$ magnetic field is {\sl self-dual (or anti-self-dual) with
respect to that metric}, $\tilde F_{mn}=\pm F_{mn}$, the 4D determinant~(\ref{det4D})
becomes a full square, hence
\bea
\sqrt{-\det\limits_{\rm 5D}\bigl(g_{MN}+2\pi\alpha'\CF_{MN}\bigr)}
&=& \sqrt{|g_{00}|}\sqrt{\det(\hat g)}\times\Bigl[
	1\,+\,(\pi\alpha')^2\hat g^{mp}\hat g^{nq}F_{mn}F_{pq}\Bigr]\nonumber\\
&=& \sqrt{|g_{00}|}\Bigl[ \sqrt{\det(\hat g)}\,
	\pm\,\tfrac12(\pi\alpha')^2\epsilon^{mnpq} F_{mn}F_{pq}\Bigr],
\eea
and the symmetrized trace becomes the ordinary
matrix trace (over the $U(2)$ gauge indices).
Consequently, the complete DBI action for all the gauge and metric fields
splits into the ordinary Yang--Mills action for the (self-dual or anti-self-dual)
$SU(2)$ magnetic fields, plus the abelian DBI action for the metric and the $U(1)$
electric fields only,
\bea
{\cal L}_{DBI} &=&
-T_8e^{-\phi}\mathop{\rm Vol}\nolimits(S^4)\sqrt{|g_{00}|}\times
	{(\pi\alpha')^2\over4}(\pm\epsilon^{mnpq}) F_{mn}^a F_{pq}^a\nonumber\\
&&\qquad-2T_8e^{-\phi}\mathop{\rm Vol}\nolimits(S^4)\times
	\sqrt{-\det_{\rm 5D}\left(g_{MN}+\pi\alpha'\hat F_{MN}\right)}
\label{DBIdecomp}
\eea
(where on the second line we have used $|g_{00}|\times\det_{\rm 4D}(\hat g_{mn})=
-\det_{\rm 5D}(g_{MN}+\pi\alpha'\hat F_{MN})$).

For the baryon, the $SU(2)$ magnetic fields become strong in the near zone, but the
$U(1)$ electric fields they induce (via the Chern--Simons interactions) are relatively
weak ($O(1/\sqrt\lambda)$ or weaker) in all zones, and the perturbations
of the 5D metric due to the scalar field $\Phi(x)$ are also weak in all zones.
Consequently, the second line of the Lagrangian~(\ref{DBIdecomp}) gives rise
to the usual kinetic terms for the scalar and the abelian gauge fields, while
all the higher-order terms are suppressed by negative powers of the
't~Hooft coupling~$\lambda$ (and outside the near zone also by positive powers
of $(\rho/r)\ll1$).
At the same time, the first line of the Lagrangian~(\ref{DBIdecomp}) gives rise
to the Yang--Mills Lagrangian for the $SU(2)$ gauge fields, and also to
their interactions with the scalar field $\Phi$ and with the background
metrics's curvature via the $u$--dependence of the 5D gauge coupling
\be
\kappa(\vec x,w)\
=\ 2(\pi\alpha')^2T_8e^{-\phi}\mathop{\rm Vol}\nolimits(S^4)\sqrt{|g_{00}|}\
=\ {N_c\over48\pi^2}\left({\bar u(w)\over\pi\alpha'}\,+\,\Phi(\vec x,w)\right).
\ee
However, we do not get any terms with higher powers of the strong $SU(2)$
gauge fields, and everything works precisely as in \S5 --- except that
the $F^a_{mn}(\vec x,w)$ fields should be self dual with respect to
the modified metric $\hat g_{mn}$ rather than the true metric $g_{mn}$.

To see the effect of this modification of the self-duality condition,
consider an $SO(4)$ spherically symmetric instanton-like field configuration
\be
A_0(\vec x,w)\ =\ 0,\quad
A_m(\vec x,w)\ =\ h(r)\times -iU^\dagger\partial_m U,\quad
U(\vec x,w)\ =\ {w+i\vec\tau\cdot\vec x\over r}\,,\quad
r\,=\,\sqrt{\vec x^2+w^2}
\label{Ilike}
\ee
where $m=1,2,3,w$ and $h(r)$ is some smooth function of the 4D radius.
The corresponding magnetic field strength has form
\be
F_{mn}(\vec x,w)\ =\ -i\partial_{[m} h\times U^\dagger\partial_{n]} U\
+\ ih(1-h)\times\left[U^\dagger\partial_m U,U^\dagger\partial_n U\right].
\label{Itension}
\ee
In particular, along the $w$ axis we have
\be
F^a_{wi}\ =\ -F^a_{iw}\ =\ {2\over r}\,{dh\over dr}\times \delta^a_i\,,\qquad
F^a_{ij}\ =\ {4h(1-h)\over r^2}\times\epsilon^{aij},
\label{Waxis}
\ee
while in other directions we have similar fields rotated by appropriate
$SO(4)\times SU(2)$ symmetries.
The instanton density of such fields is
\be
I\ =\ {\epsilon^{k\ell mn} F^a_{k\ell}F^a_{mn}\over 64\pi^2}\
=\ {3\over\pi^2}\,{h(1-h)\over r^3}\,{dh\over dr}\,,
\label{IDoB}
\ee
so the net instanton number is
\be
\#{\rm Instantons}\
=\int\limits_0^\infty\!\!I(r)\times 2\pi^2r^3\,dr\
=\int\limits_{h(0)=0}^{h(\infty)=1}\!\!6h(1-h)\,dh\ =\ 1
\ee
for any radial profile $h(r)$ that satisfies the boundary conditions.

The magnetic fields~(\ref{Itension}) become self-dual
$F^a_{mn}=\tfrac12\epsilon_{mnpq}F^a_{pq}$
{\sl with respect to the flat 4D metric}
$\delta_{mn}$ when
\be
{2\over r}\,{dh\over dr}\ =\ {4h(1-h)\over r^2}\,;
\ee
solving this differential equation gives us the usual instanton profile
\be
h_{\rm instanton}(r)\ =\ {r^2\over r^2+\rho^2}\quad
\hbox{for some \sl constant}\ \rho.
\label{instprof}
\ee
The self-duality condition with respect to the modified 4D metric $\hat g_{mn}$
calls for a slightly different profile
\be
h_{\rm mod}(r)\ =\ {r^2\over r^2+\rho^2\times(1+\delta(r))}
\label{modprof}
\ee
where the correction $\delta(r)$ depends on the abelian and the scalar fields.
We shall see momentarily that this correction is rather small, $\delta(r)=O(1/\lambda)$,
so the baryon profile~(\ref{modprof}) is approximately the usual instanton
profile~(\ref{instprof}).

To obtain the modified metric $\hat g_{mn}$ and the corresponding self-duality
condition we need the ordinary 5D metric $g_{MN}$.
Eq.~(\ref{M5D}) gives us this metric to first order in the scalar field
$\Phi$, and that's a good enough approximation even in the near zone where
\be
\Phi(\vec x,w)\ =\ O(M_\Lambda)\ \ll\ {\bar u(w)\over\alpha'}\ =\ O(\lambda M_\Lambda).
\ee
Consequently, expanding the modified metric
\be
d\hat s^2\ \equiv\ \hat g_{mn}dx^m dx^n\
=\ g_{mn}dx^m dx^n\ -\ |g^{00}|\bigl(\pi\alpha'\partial_m\hat A_0\,dx^m\bigr)^2
\ee
in powers of $1/\lambda$
(in the near zone where $\delta\bar u(w)\equiv\bar u(w)-u_0\lesssim(u_0/\lambda)$),
we get
\begin{subequations}
\label{ModG}
\begin{align}
d\hat s^2\ &
=\ \left({u_0\over R_{D4}}\right)^{3/2}\Bigl[ d\vec x^2\,+\,dw^2\,+\,O(1/\lambda)\Bigr]\\
&=\ \left({u_0\over R_{D4}}\right)^{3/2}\biggl[
	\left(1\,+\,{3\delta\bar u+3\pi\alpha'\Phi\over 2 u_0}\right)\,d\vec x^2\,
	+\,\left(1\,+\,{3\delta\bar u-13\pi\alpha'\Phi\over 2 u_0}\right)\,dw^2\\
&\hskip 7em+\,(\pi\alpha')^2\left({R_{D4}\over u_0}\right)^3\left(
	{1\over1-\zeta^{-3}}\,\bigl(\partial_m\Phi\,dx^m\bigr)^2\,
	-\,\bigl(\partial_m\hat A_0\,dx^m\bigr)^2
	\right)\quad\\
&\hskip 7em+\,O(1/\lambda^2)\biggr].
\end{align}
\end{subequations}
Note that to the zeroth order in $1/\lambda$, the modified metric (\ref{ModG}a)
of the near zone is simply flat, so the modified self-duality condition is just the
good old flat-space self-duality condition $F^a_{mn}=\tfrac12\epsilon_{mnpq}F^a_{pq}$.
And that's why the DBI correction $\delta(r)$ to the baryon profile~(\ref{modprof})
is $O(1/\lambda)$, hence
\be
R_{\rm baryon}[{\rm full\ DBI}]\ =\ R_{\rm baryon}[{\rm as\ in\ eq.}~\ref{Rbaryon}]
\times(1+O(1/\lambda)),
\ee
and all our results of \S5 for the intermediate-zone and far-zone nuclear forces
are indeed correct to leading order in $1/\lambda$.

For completeness sake, let's calculate the modified baryon profile~(\ref{modprof})
to first order in $1/\lambda$ using the modified metric~(\ref{ModG}b--d).
Since the overall conformal factor of $\hat g_{nm}$ does not affect the self-duality
condition~(\ref{MHD}), let's factor it out as
\be
{\cal C}\ =\ \left({\bar u(w)\over R_{D4}}\right)^{3/2}
\left(1\,-\,{2\pi\alpha'\Phi\over\bar u}\,+\,\cdots\right)
\label{ConFactor}
\ee
and focus on the remaining deviations of the $\hat g_{mn}$ from flatness,
\be
d\hat s^2\ =\
{\cal C}\times\biggl[
	(d\vec x^2+dw^2)\ +\ {\cal D}(r)\times dr^2\
	+\ {2\pi\alpha'\Phi\over\bar u}\times\bigl(d\vec x^2-3dw^2\bigr)\
	+\ O(1/\lambda^2)\biggr]
\label{NonConf}
\ee
where
\be
{\cal D}(r)\ =\ (\pi\alpha')^2\left({R_{D4}\over u_0}\right)^3\times
\left({(d\Phi/ dr)^2\over 1-\zeta^{-3}}\,-\,(d\hat A_0/dr)^2\right).
\label{Ddef}
\ee
Note that the first two terms inside the square brackets in (\ref{NonConf})
are spherically symmetric in all 4 space dimensions, but this
$SO(4)$ symmetry between the $\vec x$ and $w$ coordinates
is broken by the third term.
Consequently,  a single static baryon is $SO(4)$ symmetric only to the
leading order of the $1/\lambda$ expansion but the sub-leading terms
spoil this symmetry.
To see this asymmetry we should use a more complicated ansatz than~(\ref{Ilike})
for the non-abelian gauge fields, so let's leave this issue for future research.
For now let us focus on  the baryon's radial profile, and in first-order
perturbation theory this profile depends only on the spherically symmetric part
\be
d\hat s^2_{\rm symm}\ =\ {\cal C}\times\Bigl( \bigl(1+{\cal D}(r)\bigr)\times dr^2
	+\ r^2\times d\Omega_3\Bigr)
\label{sympart}
\ee
of the modified metric $\hat g_{mn}$.

For a spherically symmetric metric~(\ref{sympart}) the self-duality condition
(\ref{MHD}) becomes very simple: The $SU(2)$ magnetic fields $F_{mn}^a$
with one radial and one tangential index should be $\sqrt{1+{\cal D}}$ times stronger
than the fields with two tangential indices, for example along the $w$ axis
we should have
\be
F^a_{wi}\ =\ \sqrt{1+{\cal D}}\times\tfrac12\epsilon_{ijk}F^a_{jk}\,.
\ee
In terms of the radial profile $h(r)$ of the spherically symmetric $SU(2)$
fields~(\ref{Ilike}), the modified self-duality condition amounts to
\be
{2\over r}\,{dh\over dr}\ =\ \sqrt{1+{\cal D}(r)}\times{4h(1-h)\over r^2}\,.
\label{DSeq}
\ee
Solutions of this differential equation have general form
\be
{h(r)\over 1-h(r)}\ =\ \int\!{2dr\over r}\,\sqrt {1+{\cal D}(r)}\
+\ \rm const,
\ee
and for ${\cal D}(r)=O(1/\lambda)\ll1$ as in eq.~(\ref{Ddef}) we may approximate
these solutions as
$$
h(r)\ =\ {r^2\over r^2+\rho^2(1+\delta(r))}
\eqno(\ref{modprof})
$$
where
\be
\delta(r)\ =\int\limits^\infty_r{{\cal D}(r')\,dr'\over r'}\,.
\ee
In the first-order perturbation theory we may calculate ${\cal D}(r)$
using the scalar and electric fields of the un-perturbed baryon.
In the near zone
\bea
\hat A_0(r) &=&
{27\pi\over\lambda M_\Lambda\zeta}\times{r^2+2\rho^2\over(r^2+\rho^2)^2}\,,
\label{ANear}\\
\Phi(r)\ &=& -{1-\zeta^{-3}\over 3}\times\hat A_0(r),
\label{PhiNear}
\eea
hence
\bea
{\cal D}(r) &=&
-(1-\rar)\times\left({R_{D4}\over\ u_0}\right)^2
	\left({54\pi^2\over\lambda M_\Lambda}\right)^2\times
	{r^2(r^2+3\rho^2)^2\over(r^2+\rho^2)^6}\nonumber\\
&=& -{\pi\over \lambda}\times{(40\zeta^3-25)^{3/2}\over 2^{3/2}\zeta^5\sqrt{1-\rar}}
\times{(r/\rho)^2(3+(r/\rho)^2)^2\over4(1+(r/\rho)^2)^6}
\eea
and
\be
\delta(r)\ =\ -{\pi\over \lambda}\times
{(40\zeta^3-25)^{3/2}\over 2^{3/2}\zeta^5\sqrt{1-\rar}}\times
{32+25(r/\rho)^2+5(r/\rho)^4\over 120(1+(r/\rho)^2)^5}\,.
\ee
As  promised, this correction to the baryon profile is
small in the near zone --- $\delta(r\sim\rho)=O(1/\lambda)$ ---
and becomes even smaller at larger radii.

\medskip\centerline{$\blue\star\quad\star\quad\star$}\smallskip

Thus far, we have merely {\it assumed} the self-duality of the $SU(2)$ magnetic
fields with respect to the modified 4D metric $\hat g_{mn}$.
To justify this assumption, we are now going to show that the lowest-energy field
configuration with $N_{\rm instanton}=1$ is indeed self-dual, or at least
approximately self-dual to leading order in $1/\lambda$.

For simplicity, we look for the minimum of energy among the $SO(4)$ symmetric
field configurations only, $i.\,e.$ we restrict the $SU(2)$ gauge fields
to the ansatz~(\ref{Ilike}) but allow for generic radial profiles $h(r)$
(subject to boundary conditions $h(0)=0$, $h(\infty)=1$).
Likewise, we take the modified 4D metric $\hat g_{mn}$ to be spherically
symmetric as in eq.~(\ref{sympart}).
For such configurations, the 4D determinant~(\ref{det4D}) becomes
\begin{align}
\det\bigl(\hat g_{mn}+2\pi\alpha'F_{mn}\bigr)\,&
=\,\det(\hat g_{mn})\times\left[ 1
	+2(\pi\alpha')^2\hat g^{mp}\hat g^{nq}F_{mn}F_{pq}
	+(\pi\alpha')^4\left(\hat g^{mp}\hat g^{nq}F_{mn}\tilde F_{pq}\right)^2
	\right]\nonumber\\
&=\, {\cal C}^4(1+{\cal D})\times
	\bigl(1+\alpha^2\vec\tau^2\bigr)\bigl(1+\beta^2\vec\tau^2\bigr)
\label{BigDet}
\end{align}
where
\be
\alpha\ =\ {2\pi\alpha'\over\cal C}\times{2h(1-h)\over r^2}\,,\qquad
\beta\ =\ {2\pi\alpha'\over\cal C}\times{1\over r}\,{dh\over dr}
\times(1+{\cal D})^{-1/2},
\label{AlphaBeta}
\ee
and $\vec\tau^2=\tau_1^2+\tau_2^2+\tau_3^2$.
Please note that we are not allowed to use the Pauli matrix algebra
and set $\vec\tau^2=3$ while we calculate the determinant~(\ref{BigDet});
instead, we should to treat the $\tau_i$ as if they were independent
and un-constrained commuting numbers.

Given the determinant (\ref{BigDet}), we need to expand its square root
in powers of $\alpha\tau_i$ and $\beta\tau_i$,
then for each term in this expansion we should
restore the $SU(2)$ gauge indices of the $\tau_{1,2,3}$ factors and calculate
the symmetrized trace, and then we need to re-sum the series.
This is easy to do in the self-dual case of $\alpha=\beta$ when the determinant
(\ref{BigDet}) is a full square, but for general $\alpha\neq\beta$ this calculation
takes a few pages.
We present it in the Appendix to this article; here is the end result:
\be
\str\sqrt{(1+\alpha^2\vec\tau^2)(1+\beta^2\vec\tau^2)}\
=\ {2+4\alpha^2+4\beta^2+6\alpha^2\beta^2
	\over\sqrt{(1+\alpha^2)(1+\beta^2)}}\,.
\label{netSTR}
\ee
For a given product $\alpha\times\beta$ this expression is minimized for
self-dual $\alpha=\beta$ --- this is the DBI version of the BPS lower bound
on the Young--Mills action for gauge fields of a given instanton number.
For convenience, we may rewrite eq.~(\ref{netSTR}) as
\be
\str\sqrt{(1+\alpha^2\vec\tau^2)(1+\beta^2\vec\tau^2)}\
=\ 2\,+\,6\alpha\beta\,+\,P(\alpha,\beta)\times(\alpha-\beta)^2
\label{Pdef}
\ee
where $P(\alpha,\beta)$ is some complicated  expression.
It's gory details will not be important in the following,
expect for the special case of approximately self-dual $\alpha\approx\beta$
when
\be
P(\alpha\approx\beta)\ \approx\ {(3+\alpha\beta)\over(1+\alpha^2)(1+\beta^2)}\,.
\label{PSD}
\ee
More generally, $P(\alpha,\beta)$ is always positive and never greater
then~3.
For weak fields $\alpha,\beta\gg1$, $P\approx3$ --- indeed, the Yang--Mills
Lagrangian $\tr(1+\half(\alpha+\beta)^2\vec\tau^2)=2+3(\alpha+\beta)^2$
has form~(\ref{Pdef}) for $P=3$ --- so the difference $3-P$ measures
the importance of the higher-order DBI corrections beyond the Yang--Mills
approximation.
For example, the approximately self-dual baryon with $h\approx r^2/(r^2+\rho^2)$
for $\rho$ as in eq.~(\ref{Rbaryon}) has
\be
\alpha(r)\ \approx\ \beta(r)\
\approx\ {2\pi\alpha'\over\cal C}\times{2\rho^2\over(r^2+\rho^2)^2}\
=\ {\sqrt{k(\zeta)}\times\rho^4\over(r^2+\rho^2)^2}
\quad{\rm for}\ k(\zeta)\ =\ {40\zeta^3-25\over 64\zeta^3+8}
\label{SDABK}
\ee
and hence
\be
P(r)\ \equiv\ P(\alpha(r),\beta(r))\
=\ 3\ -\ {5k\over(1+(r/\rho)^2)^4+k}\ +\ {2k^2\over[(1+(r/\rho)^2)^4+k]^2}\,.
\label{Pfla}
\ee
At the very center of the baryon, $P(r=0)$ dips to 2.20 for the antipodal model ---
and even lower to 1.37 for the non-antipodal models with $\zeta\gg1$ ---
which indicates fairly strong higher-order DBI corrections.
However, just one unit of $\rho$ outside the center, $P(r=\rho)$ climbs back
to 2.8 for the $\zeta\gg1$ models and even higher for the antipodal model,
so the higher-order DBI corrections are important
only in the inner core $r\lesssim\rho$ of the baryon's near zone.
Outside this inner core --- from the outer side of the near zone all the way
to the far zone --- the higher-order DBI terms are small and one may
safely use the Yang--Mills approximation to the $SU(2)$ fields' Lagrangian.

Plugging eq.~(\ref{Pdef}) into the non-abelian DBI Lagrangian, we obtain
\begin{subequations}
\begin{align}
-{\cal L}_{\rm DBI}\
&=\ T_8e^{-\phi}{\rm Vol}(S^4)\sqrt{|g_{00}|}\times
\str\sqrt{\det(\hat g_{mn}+2\pi\alpha'F_{mn})}\\
&=\ {T_8{\rm Vol}(S^4)\over g_s}\times{\cal C}^2\sqrt{1+\cal D}\times
\Bigl( 2\,+\,6\alpha\beta\,+\,P(\alpha,\beta)(\alpha-\beta)^2\Bigr)\\
&=\ {T_8{\rm Vol}(S^4)\over g_s}\times
\left[\,\vcenter{\normalbaselines
  \ialign{$\displaystyle{{}#}$\hfil\cr
	2{\cal C}^2\sqrt{1+\cal D}\cr
	+\,12(2\pi\alpha')^2\times{h(1-h)\over r^3}\,{dh\over dr}\cr
	+\,(2\pi\alpha')^2 {P(\alpha,\beta)\over\sqrt{1+\cal D}}\times
		\left({1\over r}\,{dh\over dr}\,
			-\,{2h(1-h)\over r^2}\times\sqrt{1+\cal D}\right)^2\cr
	}}\right]\qquad
\label{Triple}
\end{align}
\end{subequations}
Inside the brackets on the bottom line~(\ref{Triple}) of this formula,
the first term gives rise to the DBI Lagrangian for the scalar and abelian
vector fields but does not affect the $SU(2)$ gauge fields.
The other two terms generate the Hamiltonian for
the static $SU(2)$ magnetic fields in a fixed background
of the other fields $\Phi(r)$ and $\hat A_0(r)$.
The second term provides the minimal BPS energy of the self-dual fields,
while the third term is the energy cost of deviations from self-duality:
It vanishes precisely when the modified self-duality equation~(\ref{DSeq})
is satisfied.
Adding the Chern--Simons term
\be
{\cal H}_{\rm CS}\ =\ {N_c\over2}\,\hat A_0(r)\times I(r)\
=\ {3N_c\hat A_0(r)\over 2\pi^2}\times {h(1-h)\over r^3}\,{dh\over dr}
\ee
to the Hamiltonian for magnetic $SU(2)$ fields in a fixed background
of $\hat A_0$ and $\Phi$ fields, we obtain
\bea
{\cal H} &=& \hbox{$h$--independent terms}\nonumber\\
&& +{N_c\over 2\pi^2}\left({\bar u\over\pi\alpha'}+\Phi+3\hat A_0\right)
		\times{h(1-h)\over r^3}\,{dh\over dr}\nonumber\\
&&+ {N_c\over 24\pi^2}\left({\bar u\over\pi\alpha'}+\Phi\right)\,
		{P(\alpha,\beta)\over \sqrt{1+\cal D}}\times
		\left({1\over r}\,{dh\over dr}\,
			-\,{2h(1-h)\over r^2}\times\sqrt{1+\cal D}\right)^2.
\label{Hdensity}
\eea
Using spherically averaged brane geometry
\bea
\bar u(x)\,=\, u_0\,+\,\delta\bar u(x),\nonumber\\
\delta\bar u(w\to r) &\approx&
{16\zeta^3-10\over 81\zeta}\,\lambda\alpha' M_\Lambda^3\times
\left(w^2\,\to\,{r^2\over4}\right)\\
&\approx&
M_\Lambda\alpha'\sqrt{{8\zeta^3-5\over 10\zeta^2}\times(1-\rar)}\times(r/\rho)^2
\nonumber
\eea
in the near zone, we integrate this Hamiltonian density to
\begin{subequations}
\label{BigHam}
\begin{align}
H[h(r)]\, ={}\, &
\hbox{$h$--independent}\
+\ {N_c u_0\over\pi\alpha'}\int_0^\infty\!dr\,h(1-h)\,{dh\over dr}\\
&+\, N_c\!\int\limits_0^\infty\!dr\,h(1-h)\,{dh\over dr}\times
	\left( {\delta\bar u(r)\over\pi\alpha'}\,+\,3\hat A_0(r)\,+\,\Phi(r)\right)\\
&+\, N_c\!\int\limits_0^\infty\!dr\,r\,{P(\alpha(r),\beta(r))\over\sqrt{1+\cal D}}\times
	\left({dh\over dr}\,-\,2h(1-h)\times{\sqrt{1+\cal D}\over r}\right)^2
	\times{u_0(1+O(1/\lambda))\over12\pi\alpha'}\,.\end{align}
\end{subequations}
Note that the integral on the top line~(\ref{BigHam}a) has the same value $\frac16$
for any baryon profile $h(r)$ satisfying the boundary conditions;
this integral provides the leading contribution $(\lambda N_c M_\Lambda\zeta)/27\pi$
to the baryon's mass.
The second line~(\ref{BigHam}b) corrects the BPS energy
of self-dual $SU(2)$ fields in a non-uniform background of $\hat A_0(r)$
and $\Phi(r)$ fields, and also brane curvature $\delta u(r)$.
This extra energy has a non-trivial dependence on the baryon's radius and profile;
its overall magnitude is $O(N_c M_\Lambda)=O(\lambda^{-1}M_{\rm baryon})$.
Finally, the third line~(\ref{BigHam}c) is the energy cost of deviation
from self-duality.
The $u$-dependent factor here is $u_0$ (rather than much smaller
$\delta\bar u+\pi\alpha'(3\hat A_0+\Phi)$ on the second line), so the energy
cost of a major non-self-duality would be $O(M_{\rm baryon})$.
And minimizing this extra energy is precisely why the baryon profile is
approximately self-dual.

Indeed, minimizing the Hamiltonian~(\ref{BigHam})
as a functional of the instanton profile $h(r)$
gives us a rather messy differential equation
\begin{align}
\left({d\over dr}\,+\,{2(1-2h)\over r}\,\sqrt{1+\cal D}\right)
\biggl[P(\alpha(r),\beta(r)) &
\times\left({r\over\sqrt{1+\cal D}}\,{dh\over dr}\,-\,2h(1-h)\right)\biggr]\ ={}
\label{BigMess}\\
&=\ -6h(1-h)\times{d\over dr}\,{\delta\bar u+\pi\alpha'(3\hat A_0+\Phi)\over u_0}\,.
\nonumber
\end{align}
In the near zone, the ratio $(\delta\bar u+\cdots)/u_0$ on the right hand
side is $O(1/\lambda)$ small,
so on the left hand side we should have a similarly small
\be
{r\over\sqrt{1+\cal D}}\,{dh\over dr}\,-\,2h(1-h)\
=\ O(1/\lambda)
\ee
deviation from self-duality with respect to the modified metric $\hat g_{mn}$.
The net energy cost of this deviation is $O(\lambda^{-2}M_{\rm baryon})$,
which is too small to be concerned with at the present level of analysis.
Likewise,  the effects of self-duality violation on the
baryon's average radius ---
and hence on the nuclear forces in the intermediate and far zones ---
are minor corrections of relative order $O(1/\lambda)$ to the leading-order
effects we have calculated in the previous section~5.

To conclude this section, we notice that while the deviation of the baryon'
profile from the modified self-duality condition is $O(1/\lambda)$ small,
it is comparable to the modification of flat-space self-duality condition
due to ${\cal D}(r)\neq0$.
So for completeness sake, we would like to calculate both effects on
the baryon profile to the same order $O(1/\lambda)$.
Let's parametrize deviations of $h(r)$ from a flat-space instanton
using $\delta(r)$ as in
$$
h(r)\ =\ {r^2\over r^2+\rho^2\times(1+\delta(r))}\,.
\eqno(\ref{modprof})
$$
Substituting this formula into the differential equation~(\ref{BigMess})\
and expanding to first order in $\delta(r)$ and ${\cal D}(r)$, we get
\be
\left({d\over dr}\,+\,{2(\rho^2-r^2)\over r(\rho^2+r^2)}\right)
\left[{\rho^2 r^2 P(r)\over (\rho^2+r^2)^2}\times
	\left(r{d\delta\over dr}\,+\,{\cal D}\right)\right]\
=\ {6\rho^2 r^2\over (\rho^2+r^2)^2}\times
{d\over dr}\,{\delta\bar u+\pi\alpha'(3\hat A_0+\Phi)\over u_0}\,,
\label{SmallMess}
\ee
where on the left hand side $P(r)$ is as in eqs.~(\ref{SDABK}--\ref{Pfla})
and on the right hand side
\bea
{\delta\bar u+\pi\alpha'(3\hat A_0+\Phi)\over u_0} &=&
T\left[ {r^2\over 5\rho^2}\,+\,{\rho^2(2\rho^2+r^2)\over(\rho^2+r^2)^2}\right]\\
{\rm where}\ T &=&
{9\pi\over\lambda\zeta^2}\sqrt{40\zeta^3-25\over8}\sqrt{1-\rar}\
=\ O(1/\lambda),
\eea
Solving the differential equation~(\ref{SmallMess}) is a straightforward
exercise in calculus.
Integrating the outer differential operators gives us
\be
r{d\delta\over dr}\,+\,{\cal D}\
=\ -{T\over P(r)}\times{2r^2(3r^4+12r^2\rho^2+14\rho^4)\over5\rho^2(r^2+\rho^2)^2}\,.
\label{LastStep}
\ee
Note that the right hand side here is of the same magnitude as the $\cal D$
term on the right hand side which modifies the self-duality condition;
indeed, in present notations
\be
{\cal D}\ =\ -2kT\times{\rho^6r^2(3\rho^2+r^2)^2\over(r^2+\rho^2)^6}\,.
\ee
Altogether,
\be
\delta(r)\ =\ T\!\int\!{2r\,dr\over(r^2+\rho^2)^6}\left(
	k\rho^6(3\rho^2+r^2)^2\,
	-\,{[3(r^2+\rho^2)^2+2\rho^4]\times[(\rho^2+r^2)^4+k\rho^8]^2
		\over 5\rho^2 [3(\rho^2+r^2)^4+k\rho^8]}
	\right)
\ee
where the analytic form of the integral is rather unwieldy.
Instead of writing it as a formula,
let's plot $\delta(r)$ for 3 representative values of~$\zeta$:
\be
\psset{xunit=4cm,yunit=1.5cm,shift=-3}

\ee
The differences between the three colored lines here are due to the higher-order
(beyond YM) terms in the DBI action whose effect depends on the $\zeta$ parameter.
The low-order interactions  --- Yang--Mills, Chern--Simons, $\Phi\CF^2$,
and the D8-brane curvature $\delta\bar u(r)$ ---
have the same radial dependence (in units of $\rho$) for all the Sakai--Sugimoto models,
so when the higher-order interactions pucker out at $r\gtrsim\rho$, the
deviations from self-duality become $\zeta$--independent and all the colored
lines converge to the same black-dotted line
\be
\delta_0(r)\ \approx\ {T\over 5}\left(
	-{r^2\over\rho^2}\,-\,2\log{r^2+\rho^2\over\rho^2}\,
	+\,{(5/3)\rho^2\over r^2+\rho^2}\right).
\label{SimpCurve}
\ee
This line shows that the low-order interactions themselves make the $SU(2)$
fields deviate form self duality.
In particular, at large radii  we have a growing
deviation $\delta\propto -r^2$ due to brane curvature
$\delta\bar u\propto +r^2$ (and hence the 5D gauge coupling growing with $r$).
Consequently, self-duality of the $SU(2)$ fields becomes less accurate
in the intermediate zone $r\gg\rho$, and eventually breaks down in the
far zone $r\gtrsim(\rho/\sqrt{T})\sim\rho\sqrt\lambda\sim M_\Lambda^{-1}$.

%
\section{Summary and Open Questions}
Viable holographic nuclear physics obviously requires both attractive
and repulsive  nuclear forces.
In holographic context, the hard core repulsive potential was found  in \cite{Hashimoto:2009ys},
and in this article we saw that the non-antipodal version of the Sakai--Sugimoto model
gives rise to the attractive potential as well as repulsive.
Let us summarize our main results:

\begin{itemize}
\item
We argue that  nuclear physics in the $N_c\to\infty$ limit could be quite different
from the real-life case of $N_c=3$, which limits the applicability of holography
to zero-temperature nuclear physics.
In particular, the ratio of kinetic to potential energy of the bulk nuclear matter
scales with $N_c$ as $1/N_c^2$, and consequently nuclear matter is a Fermi liquid for small
$N_c$ but becomes a crystalline solid for large $N_c$.
We estimate the transition between liquid and solid phases happens for $N_c\sim 8$,
but this estimate is rather crude and should be taken with a large grain of salt.

\item
Holographically, the attractive forces between nucleons arise from the coupling of
the gauge fields living on the flavor branes to the scalar fields parametrizing
fluctuations of the those branes' geometry;
in 5D, this coupling has form
\be
S_{\rm 5D}\ \supset\int\!\!d^5x\, \Phi\times\tr\bigl(F_{MN}^2\bigr).
\label{VS5D}
\ee
The antipodal Sakai--Sugimoto model has an accidental $\Phi\to-\Phi$ symmetry which
forbids this coupling, so in that model there are no attractive forces.
But the non-antipodal models don't have this symmetry, and consequently they do
have the scalar-vector coupling~(\ref{VS5D}) and hence the attractive nuclear forces.

\item
At intermediate distances $r$ between the two nucleons --- larger than the nucleon
radius $\rho\sim R_{KK}/\sqrt\lambda$ but smaller than the Kaluza--Klein scale $R_{KK}$,
see diagram~(\ref{zones}) on page~\pageref{zones} ---
both the attractive and the repulsive potentials have the 5D Coulomb form,
$V(r)\propto 1/r^2$.
But the attractive potential has a smaller coefficient, so the net central potential
is repulsive,
\be
\rar\ \equiv\ {-V^{\rm attractive}\over V^{\rm repulsive}}\
=\ {1-\zeta^{-3}\over 9}\ <\ \frac19\ <\ 1\quad\Longrightarrow\quad
V^{\rm net}(r)\ >\ 0.
\ee

\item
There are similar attractive forces between different parts of the same baryon,
and they reduce the baryon's radius by a factor $(1-\rar)^{1/4}$.
Consequently, the isovector spin-spin and tensor forces between two baryons are
reduced by an overall factor $(1-\rar)$.

\item
At longer distances $r\gtrsim R_{KK}\sim 1/M_{\rm meson}$, the nuclear forces are dominated
by 4D Yukawa forces due to the lightest meson with appropriate quantum numbers:
The vector isosinglet for the repulsive central forces and the scalar isosinglet
for the attractive central forces.
In the Sakai--Sugimoto model, the lightest scalar meson is heavier than the lightest
vector meson, and consequently the attractive force has a shorter range than the
repulsive force.
It is not clear whether this un-realistic behavior is peculiar to the Sakai--Sugimoto
models or a general problem of the large $N_c$ limit.
Indeed, in real life the lightest scalar meson is $\sigma(600)$ but it's QCD origin
is not clear.
If it happens to be a bound state of two pions rather than a true $q\bar q$ meson,
then in the large $N_c$ limit this bound state will fall apart and the lightest
surviving scalar meson would be heavier than the lightest vector meson $\omega(787)$.

\item
The vector fields living on the flavor branes are governed by the $\rm DBI+CS$
action, but usually the DBI part of the action is truncated to the lowest-order Yang--Mills terms,
\be
{\cal L}_{\rm DBI}\ \propto\ \str\sqrt{\det(g+2\pi\alpha'\CF)}\
\approx\ \sqrt{-\det(g)}\times\left(
	N_f\,+\,{(2\pi\alpha')^2\over4}\tr\left(\CF_{MN}\CF^{MN}\right)\right).
\ee
In the Sakai--Sugimoto model with $\lambda\gg1$, this approximation is valid at
intermediate and long distances from baryons, but inside the instanton core
of a baryon the non-abelian gauge fields become too strong to neglect the higher-order
terms such as $\tr(\CF^4)$.
\unskip\nobreak\vadjust{\allowbreak\vskip\parskip}\hskip\parfillskip\break
We argue that the self-duality of the non-abelian magnetic fields saves the day
and leads to approximate cancellation of the higher-order terms.
This was known for instantons in flat space, and we show that this is also true
for the Sakai--Sugimoto baryons; all we need to do is to slightly  modify the
self-duality condition for the non-abelian fields to account for their coupling to 
the abelian electric and scalar fields.
The effect of this modification on the baryon's radial profile is quite small:
$O(1/\lambda)$ near the baryon's center and even smaller for $r\gtrsim\rho$.
\unskip\nobreak\vadjust{\allowbreak\vskip\parskip}\hskip\parfillskip\break
We have also computed numerically the deviation of the baryon's profile from
self-duality due to curvature of the flavor branes.
The net deviation from a simple YM instanton is plotted on figure~(\ref{ThreeCurves})
on page~\pageref{ThreeCurves}.
\end{itemize}

Our work gives rise to several open questions,
the biggest of which is {\it ``What happens to the $\sigma(600)$ meson
in the large $N_c$ limit?''.}
The best  answer for this question would be a lattice calculation of $m_\sigma$
for several values of $N_c$, although it's not clear if such a calculation is
possible with-present-day lattice sizes.
(But thanks to Moore's law, it should be possible in a few years.)
Alternatively, we can try several different models of holographic QCD and compare their
predictions for the lightest scalar to lightest vector mass ratio.
As of this writing, all known models have ratios${}>1$, with one exception
\cite{Dymarsky:2010ci} --- but in that model, the lightest scalar meson is
is a pseudo-Goldstone boson and probably
does not couple to the other particles like the real $\sigma(600)$ meson. 
If future holographic models show the same pattern --- the lightest true scalar meson is
either heavier than the lightest vector or else is a pseudo-Goldstone boson whose
couplings are suppressed --- then most likely, in the  large $N_c$ limit of QCD
there is no sigma meson and the attractive nuclear force has a shorter range than
the repulsive force.
But if we see a wide variation of the lightest-scalar-to-lightest-vector mass ratios
between different models, then we wouldn't know what really happens  in large-$N_c$ QCD,
but on the other hand, a holographic model with  $m_\sigma<m_\omega$  {\it might} also
have a semi-realistic nuclear potential --- repulsive at shorter distances but
attractive at longer distances.

We expect different holographic models to have different attractive / repulsive
force rations at intermediate and short distances, and it would be interesting
to see if any model has $\rar>1$.
In such a model, net attraction between different parts of the same baryon
would make it collapse to a singular point.
Or rather, a classical baryon would collapse to a point, but quantum corrections
would keep its size finite, perhaps $O(R_{KK}/\lambda)$, but much smaller
than in the Sakai--Sugimoto model.
Consequently, the net force between two such baryons would be repulsive at very
short distances $r\lesssim\rho\sim R_{KK}/\lambda$ but attractive at
intermediate distances $\rho\ll r\ll R_{KK}$.
At longer distances $r\gtrsim R_{KK}$, the net force could be either attractive
or repulsive, depending on the meson spectrum.

Another open question concerns the dependence of the effective 5D action
on the 't~Hooft coupling $\lambda$ in the effective 5D action.
In the Sakai--Sugimoto model, the flavor gauge $\rm coupling^2\propto1/\lambda$,
but this power of $\lambda$ could be different in other holographic models.
It would be interesting to find models where the flavor physics does not depend
on $\lambda$ at all and to explore the baryons and the nuclear forces in such models.
In particular, we would like to see if such  models have largish baryon radii,
$\rho\sim1/M_{\rm meson}$ like in real life, rather than $\rho\ll1/M_{\rm meson}$
as in the Sakai--Sugimoto model.
For such large-radius baryons there would be no intermediate zone of distances;
instead, the near zone (where two baryons overlap) would connect directly to the far
zone dominated by 4D Yukawa forces.
Consequently, for $r\sim\rho$ both the baryon overlap
and the curvature of the fifth dimension would be important, and the nuclear forces
in this regime would be quite different from anything in the Sakai--Sugimoto model.

There are also more general issues concerning meson spectra in holographic models.
Apart from the specific mass ratios that are probably model dependent, there are
general differences from the real-life mesons found in the {\it Particle Data Book.}
For example, the 5D scalar fields give rise to both scalar and pseudoscalar mesons
in 4D (depending on the mode number in the fifth dimension), and their charge conjugation
signs follow from parity, C-positive scalars and C-negative pseudoscalars.
But in real life, all pseudoscalar mesons are C-positive rather than C-negative.

Also, the high-spin  mesons in holographic models have
different physical origin from the low-spin  mesons and consequently
much larger masses.
While the $J=0$ and $J=1$ 4D mesons are modes of 5D scalar or vector fields,
the $J\ge2$ mesons are semi-classical rotating open strings
which start and end on flavor branes but at different points in  space
\cite{Peeters:2005fq}.
But in real life, both low-spin and high-spin mesons belong to the same Regge trajectories
\be
M^2\ =\ \alpha'\times J\ +\ \rm const
\ee
and there are no essential differences between them.

We don't know what makes the holographic meson spectra so different from the
real life, but it's almost certainly {\it not} the large $N_c$ limit.
It would be interesting to see if this problem is common to all holographic models
--- perhaps because it's inherent in the $\lambda\to\infty$ limit ---
or if there are some model with more realistic meson spectra.
If we can find such a model, maybe it would also have a light scalar meson and
hence attractive net nuclear force at longer distances.

Yet another open question concerns nuclear forces stemming from double-meson exchanges,
especially the double-pion exchange which produces a long-range attractive force.
In holography, the double meson exchanges happen at the one-string-loop level
while single meson exchanges happen at the tree level.
This makes the double exchanges smaller by a factor $1/\lambda$, and also much harder
to calculate.
But if such calculation is feasible for some holographic model, it would be very interesting
to compare its result to the real-life nuclear force due to double-pion exchanges.

Finally, there is a long-standing open problem concerning sensitivity of nuclear forces ---
and hence of the nuclear binding energy --- to the pion's mass.
Hopefully, holography can shed some new light on this old problem.
Although holographic models usually have coincident flavor branes and hence
zero current quark masses and massless pions, there are ways
\cite{Casero:2007ae,Bergman:2007pm,Dhar:2007bz,Aharony:2008an}
to explicitly break the chiral symmetry and give the pions a small mass.
It would be interesting to see if a small but non-zero $m_\pi^2$ would affect
the isoscalar central force between two nucleons, and whether such effect would
happen at the tree level of string theory or only at the loop levels.

%
\section*{Acknowledgments}
The authors would like to thank Ofer Aharony, Jacques Distler,
Shigenori Seki, and Shimon Yankielowicz for many fruitful conversations.
We also thank Alexei Cherman and Tom Cohen for explaining to us how to count
the powers of $N_c$ in nuclear forces involving multiple meson exchanges.

The research presented in this article was supported by:
The US--Israel Binational Science Foundation (both authors),
the US National Science Foundation (V.~K., grant \#PHY--0455649),
the Israel Science Foundation (J.~S., grant\#1468/06),
the German--Israeli Project Cooperation (J.~S., grant\#DIP~H52),
and German--Israeli Foundation (J.~S.).

%
\appendix
\section{Symmetrized Trace of the Non-Abelian DBI Action}
According to Tseytlin \cite{Tseytlin:1997csa}, the non-abelian version
of the Dirac--Born--Infeld Lagrangian $\sqrt{\det(g_{MN}+2\pi\alpha'\CF_{MN})}$
works like this:
First we focus on the spacetime indices of the $g_{MN}+2\pi\alpha'\CF_{MN}$
and formally calculate the determinant of this $d\times d$ matrix
while completely ignoring the gauge indices of the $\CF_{MN}$ fields
or the fact that they don't commute with each other.
In other words, at this stage of the calculation, we treat each
component $\CF_{MN}$ as if it was a just real number rather than a generator of
some non-abelian group.
Second, we expand the square root of the determinant into a power series in
the $\CF_{MN}$ fields; again, we ignore the fields' non-commutativity and treat them
as real numbers.
Third, for each term in the expansion, we restore the gauge indices of the fields
(in the fundamental representation of a $U(N)$ group), symmetrize the product
of non-commuting fields, and take the trace,
\be
\str\Bigl(\CF_{M_1N_1}\CF_{M_2N_2}\cdots\CF_{M_kN_k}\Bigr)
=\ {1\over k!}\sum\tr\Bigl(\mbox{all permutations of}\
	\CF_{M_1N_1}\CF_{M_2N_2}\cdots\CF_{M_kN_k}\Bigr).
\ee
Finally, we try to re-sum the power series in the $\CF$ fields;
if we a lucky, it might have a nice analytic form.

In section~6, we had  spherically-symmetric (in 4D)
$SU(2)$ fields~(\ref{Itension}), and we had calculated the DBI
determinant for those fields as
$$
\det\Bigl(\hat g_{mn}+2\pi\alpha'\CF_{mn}\Bigr)\
=\det(\hat g_{mn})\times\left(1+\alpha^2\vec\tau^2\right)\left(1+\beta^2\vec\tau^2\right).
\eqno(\ref{BigDet})
$$
In this Appendix, we calculate the symmetrized trace of the square root of this
determinant and show that
$$
\str\sqrt{\left(1+\alpha^2\vec\tau^2\right)\left(1+\beta^2\vec\tau^2\right)}\
=\ {2+4\alpha^2+4\beta^2+6\alpha^2\beta^2\over\sqrt{(1+\alpha^2)(1+\beta^2)}}\,.
\eqno(\ref{netSTR})
$$
Clearly, expanding the square root on the LHS into powers of $\alpha$ and $\beta$
produces all powers of $\vec\tau^2=\tau_i\tau_i$,
so our first step is to symmetrize the product $(\vec\tau^2)^n$ with respect to
all distinct permutations of the $2n$ Pauli matrices.
\par\noindent
{\bf Lemma:}
\bea
\left[\bigl(\vec\tau^2\bigr)^n\right]_{\rm symm}
&\buildrel{\rm def}\over=&
{1\over(2n-1)!!}\sum^{(2n-1)!!}\mbox{distinct permutations of}\
	\tau_{i_1}\tau_{i_1}\tau_{i_2}\tau_{i_2}\cdots\tau_{i_n}\tau_{i_n}\nonumber\\
\noalign{\vskip 10pt}
&=& (2n+1)\times{\bf 1}_{2\times2}
\label{Lemma}
\eea
where
\be
(2n-1)!!\ =\ {(2n)!\over 2^n\,n!}
\ee
is the number of distinct permutations of $2n$ matrices that come in $n$
identical pairs.
For the purpose of symmetrization, $\tau_i$ and $\tau_j$ carrying different
isovector indices $i$ and $j$ count as distinct; the summation over
$i,j,\ldots=1,2,3$ is done after the symmetrization.

The lemma~(\ref{Lemma}) is trivially true for $n=0$ and $n=1$;
indeed, for $n=1$ there is nothing to symmetrize and
$\bigl[\tau_i\tau_i\bigr]_{\rm symm}=\tau_i\tau_i=\delta_{ii}\times{\bf1}
=3\times{\bf1}$.
For the first non-trivial case $n=2$, the lemma works according to
\begin{align}
\begin{split}
\left[\bigl(\vec\tau^2\bigr)^2\,=\,\tau_i\tau_i\tau_j\tau_j\right]_{\rm symm}\ &
=\ \frac13\Bigl( \tau_i\tau_i\tau_j\tau_j\,+\,\tau_i\tau_j\tau_i\tau_j\,
	+\,\tau_j\tau_i\tau_i\tau_j\Bigr)\\
&=\ \frac13\Bigl( (\tau_i\tau_i)^2\,+\,\{\tau_i,\tau_j\}\times\tau_i\tau_j\Bigr)\\
&=\ \frac13\Bigl( (3)^2\,+\,2\delta_{ij}\times\tau_i\tau_j\
	=\ 9\,+\,2\times3\ =\ 15\Bigr)\\
&=\ 5\quad (i.\,e.,\ 5\times{\bf1}).
\end{split}
\end{align}
For higher $n>2$ the simplest proof of the lemma is recursive.
Let's group the $(2n-1)!!$ distinct permutations of the $2n$ matrices
$\tau_i\tau_i\tau_j\tau_j\cdots$ into two sets according to the two
left-most matrices being similar or distinct:
$(2n-3)!!$ of the permutations start with $\tau_i\tau_i$ (for the same $i$)
while the remaining $(2n-2)\times(2n-3)!!$ permutations start with
$\tau_i\tau_j$ --- or equivalently $\tau_j\tau_i$ --- with distinct $i$ and $j$.
Consequently,
\begin{align}
\begin{split}
(2n-1)!!\times\left[\bigl(\vec\tau^2\bigr)^n\right]_{\rm symm}\ &
=\,\sum\mbox{all distinct permutations of}\
	\tau_i\tau_i\tau_j\tau_j\tau_k\tau_k\cdots\\
&=\ \tau_i\tau_i\times\sum^{(2n-3)!!}\mbox{permutations of}\
	\tau_j\tau_j\tau_k\tau_k\cdots\\
&\qquad+\ \tfrac12\{\tau_i,\tau_j\}\times\sum^{(2n-2)(2n-3)!!}\mbox{permutations of}\
	\tau_i\tau_j\tau_k\tau_k\tau_\ell\tau_\ell\cdots\\
&=\ 3\times(2n-3)!!\left[\bigl(\vec\tau^2\bigr)^{n-1}\right]_{\rm symm}\\
&\qquad+\ \delta_{ij}\times (2n-2)(2n-3)!!
	\left[\tau_i\tau_j\bigl(\vec\tau^2\bigr)^{n-2}\right]_{\rm symm}\\
&=\ 3(2n-3)!!\times\left[\bigl(\vec\tau^2\bigr)^{n-1}\right]_{\rm symm}\\
&\qquad+\ (2n-2)(2n-3)!!\times\left[\bigl(\vec\tau^2\bigr)^{n-1}\right]_{\rm symm}\\
&=\ (3+2n-2)\times
	(2n-3)!!\left[\bigl(\vec\tau^2\bigr)^{n-1}\right]_{\rm symm}
\end{split}
\end{align}
and hence
\be
\left[\bigl(\vec\tau^2\bigr)^n\right]_{\rm symm}\
=\ \left[\bigl(\vec\tau^2\bigr)^{n-1}\right]_{\rm symm}\times\left(
	{(3+2n-2)\times(2n-3)!!\over(2n-1)!!}\,=\,{2n+1\over2n-1}\right).
\ee
Applying this formula recursively, we get
\begin{align}
\begin{split}
\left[\bigl(\vec\tau^2\bigr)^n\right]_{\rm symm}\ &
=\ {2n+1\over2n-1}\times\left[\bigl(\vec\tau^2\bigr)^{n-1}\right]_{\rm symm}\
=\ {2n+1\over2n-1}\times{2n-1\over2n-3}\times
	\left[\bigl(\vec\tau^2\bigr)^{n-2}\right]_{\rm symm}\\
\noalign{\vskip 5pt}
&=\ {2n+1\over2n-1}\times{2n-1\over2n-3}\times\cdots{5\over3}\times
	\left[\bigl(\vec\tau^2\bigr)^{1}\right]_{\rm symm}\\
\noalign{\vskip 5pt}
&=\ {2n+1\over3}\times3\times{\bf1}\ =\ (2n+1)\times{\bf1}.
\end{split}
\end{align}
{\it Quod erat demonstrandum.}

In terms of symmetrized traces, the Lemma tells us that
\be
\str\left[\bigl(\vec\tau^2\bigr)^n\right]\
\equiv\ \tr\left(\left[\bigl(\vec\tau^2\bigr)^n\right]_{\rm symm}\right)\
=\ 2(2n+1),
\ee
which leads us to the following
\par\noindent
{\bf Theorem: \it the symmetrized trace of any analytic function}
\be
f\bigl(\vec\tau^2\bigr)\ =\,\sum_{n=0}^\infty C_n\bigl(\vec\tau^2\bigr)^n
\ee
{\it of $\vec\tau^2$ can be evaluated as}
\be
\str\left[f\bigl(\vec\tau^2\bigr)\right]\
=\,\sum_n C_n\times\Bigl(\str\left[\bigl(\vec\tau^2\bigr)^n\right]\,=2(2n+1)\Bigr)\
=\ \left.\left(4x{\partial\over\partial x}\,+\,2\right) f(x)\right|_{x=1}\,.
\ee
In particular, the square root of the DBI determinant~(\ref{BigDet}) has
symmetrized trace
\begin{align}
\str\sqrt{(1+\alpha^2\vec\tau^2)(1+\beta^2\vec\tau^2)}\ &
=\ \left(4\alpha^2{\partial\over\partial\alpha^2}\,
	+\,4\beta^2{\partial\over\partial\beta^2}\,
	+\,2\right)
	\sqrt{(1+\alpha^2)(1+\beta^2)}\nonumber\\
\noalign{\vskip 5pt}
&=\ {2+4\alpha^2+4\beta^2+6\alpha^2\beta^2\over\sqrt{(1+\alpha^2)(1+\beta^2)}}\,,
\label{Answer}
\end{align}
as promised in eq.~(\ref{netSTR}).

To conclude this appendix, we note that the symmetrized trace~(\ref{Answer})
is bounded from above by the low-order tension+Yang--Mills limit $2+3(\alpha^2+\beta^2)$
and from below by its value $2+6\alpha\beta$ for the self-dual fields, thus
\be
\forall \alpha,\beta\ge0,\quad 2\,+\,6\alpha\beta\
\le\ {2+4\alpha^2+4\beta^2+6\alpha^2\beta^2\over\sqrt{(1+\alpha^2)(1+\beta)^2}}\
\le\ 2\,+\,3(\alpha^2+\beta^2).
\label{Bounds}
\ee
In terms of eq.~(\ref{Pdef}),
these bounds amount to limits on $P(\alpha,\beta)$,
\be
\str\sqrt{(1+\alpha^2\vec\tau^2)(1+\beta^2\vec\tau^2)}\
=\ 2\,+\,6\alpha\beta\,+\,P(\alpha,\beta)\times(\alpha-\beta)^2,\qquad
0\le P(\alpha,\beta)\le3.\quad
\ee

\newpage
%

\end{document}